\documentclass[a4paper,11pt]{article}
\pdfoutput=1 

\usepackage{jheppub} 
\usepackage{bbm,bm,graphicx,mathtools,color,hyperref,slashed}
\usepackage{amssymb,amsmath}

\usepackage[all]{xy}
 \usepackage[caption=false]{subfig}

\def\be{\begin{equation}}
\def\ee{\end{equation}}
\def\bea{\begin{eqnarray}}
\def\eea{\end{eqnarray}}
\def\bequ{\begin{equation}}
\def\eequ{\end{equation}}

\def\del{\partial}

\newcommand{\e}{\mathrm{e}}
\newcommand{\beq}{\begin{eqnarray}}
\newcommand{\eeq}{\end{eqnarray}}
\newcommand{\eq}{equation}
\newcommand{\eqa}{eqnarray}

\newcommand{\diff}{\mathrm{d}}
\newcommand{\p}{\partial}
\newcommand{\ve}{\varepsilon}
\newcommand{\im}{\mathrm{i}}
\newcommand{\Diff}{{\mathcal{D}}}
\newcommand{\rme}{\mathrm{e}}

\usepackage{qcircuit}
\usepackage{mathtools}

\DeclarePairedDelimiter\ket{\lvert}{\rangle}
\DeclarePairedDelimiterX\braket[2]{\langle}{\rangle}{#1 \delimsize\vert #2}

\title{
Negative string tension of higher-charge Schwinger model via digital quantum simulation}

\author[1]{Masazumi Honda,}
\affiliation[1]{Yukawa Institute for Theoretical Physics, Kyoto University, Sakyo-ku, Kyoto 606-8502, Japan}
\emailAdd{masazumi.honda(at)yukawa.kyoto-u.ac.jp}

\author[2,3,4,5]{Etsuko Itou,}
\affiliation[2]{Strangeness Nuclear Physics Laboratory,
RIKEN Nishina Center, Wako 351-0198, Japan}
\affiliation[3]{Interdisciplinary Theoretical and Mathematical Sciences Program (iTHEMS), RIKEN, Wako 351-0198, Japan}
\affiliation[4]{Department of Physics, and Research and 
Education Center for Natural Sciences, Keio University, 4-1-1 Hiyoshi, Yokohama, Kanagawa 223-8521, Japan}
\affiliation[5]{Research Center for Nuclear Physics (RCNP), Osaka University, Osaka 567-0047, Japan}
\emailAdd{itou(at)yukawa.kyoto-u.ac.jp}

\author[6]{Yuta Kikuchi,}
\affiliation[6]{Department of Physics, Brookhaven National Laboratory, Upton, NY 11973, USA}
\emailAdd{ykikuchi(at)bnl.gov}

\author[1]{Yuya Tanizaki}
\emailAdd{yuya.tanizaki(at)yukawa.kyoto-u.ac.jp}

\abstract{
We study some properties of generalized global symmetry 
for the charge-$q$ Schwinger model in the Hamiltonian formalism,
which is the $(1+1)$-dimensional quantum electrodynamics 
with a charge-$q$ Dirac fermion. 
This model has the $\mathbb{Z}_q$ $1$-form symmetry,
which is a remnant of the electric $U(1)$ $1$-form symmetry in the pure Maxwell theory.
It is known that
if we put the theory on closed space,
then the Hilbert space is decomposed into $q$ distinct sectors, called universes
and some states with higher energy density do not decay to the ground state due to the selection rule of the $1$-form symmetry. 
Even with open boundaries, 
we can observe the stability of such states 
by seeing a negative string tension behavior, meaning that opposite charges repel with each other.
We develop a method based on the adiabatic state preparation to see this feature with digital quantum simulation 
and confirm it using a classical simulator of quantum devices. 
Especially, we measure local energy density and
see how it jumps between inside and outside of insertion of the probe charges.
We explicitly see that the energy density inside is lower than the outside one.
This is a clear signature of the negative string tension. 
}
\preprint{YITP-21-111, RIKEN-iTHEMS-Report-21}

\begin{document}
\maketitle

%\clearpage
%%%%%%%%%%%%%%%%
%%%%%%%%%%%%%%%%
%%%%%%%%%%%%%%%%
\section{Introduction}
%%%%%%%%%%%%%%%%
%%%%%%%%%%%%%%%%
%%%%%%%%%%%%%%%%
Quantum field theory (QFT) is the fundamental framework to study both particle physics and quantum many-body physics. 
For weakly-coupled theories, 
we can perform perturbative expansion to compute physical quantities with sufficiently good accuracy, while  strongly-coupled QFTs are still very far from complete understanding. 
To understand properties of strongly-coupled theories, we typically take the following two approaches, or their combinations if various methods work nicely. 
One way is to constrain their possible properties from various consistencies, such as unitarity and locality, symmetry, anomaly, monotonicity of renormalization group, and so on. 
The other way is to perform the numerical computation by putting QFTs on computers 
and the most typical one would be the lattice Monte Carlo simulation. 
Every technique has its pros and cons, and we must choose an appropriate one for each problem of our interest. 
Obviously, we need to develop new techniques to extend applicability and usefulness of QFTs.

Quantum simulation is one of such promising techniques. 
A universal quantum computer provides us with a potential way to efficiently simulate the quantum system that is intractable with classical hardware~\cite{Feynman:1981tf}, and its application to QFTs may uncover new aspects of strongly-coupled many-body phenomena.\footnote{
See \cite{Jordan:2011ne,Jordan:2011ci,Jordan:2014tma,Garcia-Alvarez:2014uda,Wiese:2014rla,Marcos:2014lda,Mezzacapo:2015bra,Macridin:2018gdw,Lamm:2018siq,Klco:2018zqz,Gustafson:2019mpk,Alexandru:2019ozf,Klco:2019xro,Klco:2019evd,Lamm:2019uyc,Mueller:2019qqj,Gustafson:2019vsd,Martinez:2016yna,Muschik:2016tws,Klco:2018kyo,Kokail:2018eiw,Magnifico:2019kyj,Chakraborty:2020uhf,Yamamoto:2021vxp,Honda:2021aum} for digital quantum simulations and \cite{Zohar:2012ay,Banerjee:2012pg,Zohar:2012xf,Banerjee:2012xg,Wiese:2013uua,Zohar:2015hwa,Bazavov:2015kka,Zohar:2016iic,Bermudez:2017yrq,Zache:2018jbt,Zhang:2018ufj,Lu:2018pjk,Roy:2020ppa,Bernien_2017,Surace_2020} for analogue quantum simulations of QFTs.
}
The rapid advance in the development of quantum computing hardware, which we have witnessed recently, further motivates us to design strategies to tackle these classically hard problems. 
For example, when the Boltzmann weight of the path integral has complex phases, 
the lattice Monte Carlo simulation encounters the sign problem and 
we cannot efficiently simulate such QFTs with classical algorithm. 
This problem always occurs when we study real-time dynamics using path integral formulations. 
Also, the imaginary-time path integral can encounter the sign problem for many interesting setups, and 
revealing static properties of such a QFT is an important task. 

While a tremendous amount of efforts have been made to develop numerical techniques to tackle those problems with classical computers, their quantum counterparts are far less explored.
This is partly because the Hamiltonian formalism is more suitable for quantum simulations, but this is not the current mainstream for the study of nonperturbative QFTs. 
Instead of treating the Hilbert space directly, we usually consider the Euclidean correlation functions using the path-integral formalism. 
Especially, there is a huge development about generalized symmetries as a formal aspect of Euclidean QFTs~\cite{Pantev:2005rh, Pantev:2005wj, Pantev:2005zs, Hellerman:2006zs, Hellerman:2010fv, Gaiotto:2014kfa}, which clarifies the existence of unconventional selection rules. 
One of our motivations in this study is to decode these features of Euclidean QFTs in the Hamiltonian formalism, so that those notions can be used also in future quantum simulations.

As a first step, we study $(1+1)$-dimensional quantum electrodynamics with a charge-$q$ electron,
to which we refer as charge-$q$ Schwinger model.
This theory has $\mathbb{Z}_q$ $1$-form symmetry, denoted as $\mathbb{Z}_q^{[1]}$, and it is one of the simplest models that enjoy generalized symmetries~\cite{Anber:2018jdf, Anber:2018xek, Armoni:2018bga, Misumi:2019dwq}. 
Furthermore, when we take the fermion mass to be zero, this system has the $\mathbb{Z}_q$ chiral symmetry, and there is a mixed 't~Hooft anomaly between $\mathbb{Z}_q^{[1]}$ and $(\mathbb{Z}_q)_{\mathrm{chiral}}$. 
This shows a nontrivial commutation relation between the Wilson loop and the chiral operator, and anomaly matching requires the existence of $q$ degenerate vacua as long as the system is gapped. 
Since the massless Schwinger model can be solved exactly, we can confirm these features by explicit computations: The system is gapped due to the axial anomaly~\cite{Schwinger:1962tp, Schwinger:1962tn}, and the partition function has been computed on various two-dimensional manifolds~\cite{Manton:1985jm, Hetrick:1988yg, Sachs:1991en}. 

When the mass term is added, the system is no longer exactly solvable, 
but we can apply the mass perturbation to compute physical quantities. 
It is expected to be valid if the fermion mass is smaller than the photon mass. 
Since it breaks $(\mathbb{Z}_q)_{\mathrm{chiral}}$ explicitly, $q$ degenerate vacua is lifted, and generically we obtain the unique ground state. 
Then, let us ask the following question:
What would be the fate of other vacua? 
They are stable within the mass perturbation, but they have larger energy density compared with the ground state. 
Without having any selection rules, it would be natural to guess that they eventually experience the vacuum decay to the ground state. 

This naive guess turns out to be incorrect, and there is indeed an unconventional selection rule due to $\mathbb{Z}_q^{[1]}$. 
In general, if $d$-dimensional QFTs enjoy $(d-1)$-form symmetry, the Hilbert space completely is decomposed by the eigenvalues of the $(d-1)$-form symmetry generator~\cite{Tanizaki:2019rbk, Nguyen:2021naa}, and our example is a special case with $d=2$. 
This selection rule is stronger than the superselection rule associated with spontaneous symmetry breaking, since it is true without infinite volume limit. 
This property is called as decomposition of QFTs~\cite{Pantev:2005rh, Pantev:2005wj, Pantev:2005zs, Hellerman:2006zs, Hellerman:2010fv}, or recently dubbed as  universes~\cite{Tanizaki:2019rbk, Nguyen:2021naa, Komargodski:2020mxz, Cherman:2020cvw}: A state in one of the universes cannot jump/decay to another universe. 
In other words, any dynamical processes have to be closed within one universe. 

We can check the existence of such an exotic state by observing the negative string tension. 
String tension defines the slope of linear confinement potential of electric test particles, and it is usually positive since the confining string costs some energies. 
In $2$d QFTs, however, the Wilson loop completely separates the spacetime into two regions. 
If the universe inside the Wilson loop has lower energy density than the outside, the string tension becomes negative. 
Depending on the vacuum angle, it has been suggested that such negative string tension indeed appears~\cite{Misumi:2019dwq, Tanizaki:2019rbk}. 

In this paper, we consider the charge-$q$ Schwinger model on the open interval 
and relate it to a spin chain by Jordan-Wigner transformation. 
We prepare the ground state with test electric charges at nonzero vacuum angles, which is expected to have the negative string tension, by the adiabatic state preparation. 
In order to detect the negative string tension explicitly, it is useful to have the local energy density operator and we measure the position-dependent energy density to confirm theoretical expectations. 
We shall show that our observation about the Hilbert space is consistent with the selection rule of the generalized symmetry, $\mathbb{Z}_q^{[1]}$. 

This paper is organized as follows.
In sec.~\ref{sec:continuum}, we review the charge-$q$ Schwinger model in the continuum formulation.
In sec.~\ref{sec:lattice}, we write down the lattice formulation of the charge-$q$ Schwinger model.
In sec.~\ref{sec:strategy}, we describe our simulation strategy.
Sec.~\ref{sec:results} shows our simulation results.
Sec.~\ref{sec:discussion} is devoted to summary and discussion.
In app.~\ref{sec:2dMaxwellTheory}, we discuss these features for the $2$d pure Maxwell theory by explicit computations of the path integral and the canonical quantization.
In app.~\ref{sec:Ad-fn},
we investigate how the adiabatic schedule affects the adiabatic error.

%%%%%%%%%%%%%%%%
%%%%%%%%%%%%%%%%
%%%%%%%%%%%%%%%%
\section{Continuum theory of charge-$q$ Schwinger model}
\label{sec:continuum}
%%%%%%%%%%%%%%%%
%%%%%%%%%%%%%%%%
%%%%%%%%%%%%%%%%
In this section, we give a review on the charge-$q$ Schwinger model in the continuum formulation. 
This theory has the $\mathbb{Z}_q$ $1$-form symmetry~\cite{Pantev:2005rh, Pantev:2005wj, Pantev:2005zs, Hellerman:2006zs, Hellerman:2010fv, Gaiotto:2014kfa}, 
and this leads to the fact that the Hilbert space on closed space 
decomposes into $q$ distinct universes~\cite{Tanizaki:2019rbk}. 
In Appendix~\ref{sec:2dMaxwellTheory}, we discuss these features for the $2$d pure Maxwell theory by explicit computations of the path integral and the canonical quantization.

%%%%%%%%%%%%%%%%
%%%%%%%%%%%%%%%%
\subsection{Charge-$q$ Schwinger model on $S^1_L$ and $\mathbb{Z}_q$ $1$-form symmetry}
\label{sec:chargeq_Schwinger_1form}
%%%%%%%%%%%%%%%%
%%%%%%%%%%%%%%%%
The Lagrangian density of the charge-$q$ Schwinger model is given as\footnote{
We take the gamma matrices as $\gamma^0 =\sigma^3$, $\gamma^1 =i\sigma^2$ and $\gamma^3 =\sigma^1$.
} 
\be
\mathcal{L}
={1\over 2g^2}F_{01}^2+{\theta_0\over 2\pi}F_{01}+\overline{\psi}\,\im\, \gamma^\mu (\p_\mu + \im\, q\, A_{\mu})\psi-m\,\overline{\psi}\psi,
\ee
where 
$q$ is a positive integer and we take $m>0$.
Here, $F=\diff A$ ($F_{01}=\p_0 A_1-\p_1 A_0$ in components)
is the field strength of the $U(1)$ gauge field $A$, and we normalize it so that the Dirac quantization is given by $\int_{M_2}\diff A\in 2\pi \mathbb{Z}$ for any closed 2-manifolds $M_2$. 
Equivalently, $U(1)$ gauge transformation is imposed by the invariance under 
\be
A\mapsto A-\im\, \rme^{-\im \lambda}\,\diff\, \rme^{\im \lambda}=A+\diff \lambda,
\ee
where the gauge transformation parameter  $\lambda$ is a $2\pi$-periodic compact scalar field.  
When $\lambda$ is well-defined as $\mathbb{R}$-valued functions, 
we call it as the small gauge transformation, and others are called as the large gauge transformation. 

We first take the space as $S^1$ for the canonical quantization. 
For Hamiltonian formulation, the temporal gauge $A_0=0$ is convenient. 
When imposing this condition, we must also impose 
\be
{\delta L\over \delta A_0(x)}=0, 
\ee
which gives the Gauss law constraint,
\be
\p_1 F_{01}= q \, \psi^\dagger \psi(x) . 
\label{eq:Gauss_law_classical}
\ee
The canonical momentum for $A_1$ is 
\be
\Pi={\delta L\over \delta \dot{A_1}}={1\over g^2}\dot{A_1}+{\theta\over 2\pi}. 
\label{eq:conjugate_momentum}
\ee
Therefore the Hamiltonian density becomes
\begin{\eq}
H(x)
=\frac{g^2}{2} \left( \Pi -\frac{\theta_0}{2\pi} \right)^2
- \bar{\psi}\,\im\, \gamma^1 (\del_1 +\im\,q A_1) \psi +m\bar{\psi} \psi \label{eq:hamiltonian-cont}
.
\end{\eq}
With the canonical variables, the Gauss law constraint (\ref{eq:Gauss_law_classical}) is rewritten as 
\be 
\p_1 \Pi(x)=q \, \psi^\dagger \psi(x). 
\label{eq:Gauss_continuum}
\ee
Furthermore, we also impose the invariance under the large gauge transformation on the physical Hilbert space.

Because of the presence of the charge-$q$ electric matter, 
the $U(1)$ $1$-form symmetry in the pure Maxwell theory is explicitly broken. 
Still, this model has $\mathbb{Z}_q$ $1$-form symmetry, generated by 
\be 
A\mapsto A+\Lambda,\quad 
\psi(x)\mapsto \rme^{-\im q \int_0^x \Lambda}\psi(x),
\ee
with $\diff \Lambda=0$ and $\oint \Lambda\in {2\pi\over q}\mathbb{Z}$. 
This quantization of $\oint \Lambda$ is important to have the single-valuedness for $\psi$ after the transformation. 
For example, we can choose 
\be 
\Lambda={2\pi\over qL}\diff x ,
\ee
as a nontrivial generator for $\mathbb{Z}_q^{[1]}$. Note that $q\Lambda =\diff \phi$, where $\phi={2\pi\over L}x$ is a $2\pi$-periodic scalar, and thus it is a part of the large gauge transformation, which explains why this is $\mathbb{Z}_q$ transformation. 

The generator of $\mathbb{Z}_q^{[1]}$ is given by
\be 
U(x)=\exp\left({2\pi\im\over q}\Pi(x)\right). 
\ee
Therefore, $U(x)$ does not depend on $x$ for physical states. To see this, let $\Psi$ be a physical state which satisfies the Gauss law. Then, 
\be
U(x) U(y)^{-1} \Psi 
= \exp\left({2\pi \im\over q} \cdot q\int_y^x \diff x'\, \psi^\dagger \psi(x')\right) \Psi
%\nonumber\\
=\Psi. 
\ee
Moreover, $U(x)$ commutes with the Hamiltonian. The nontrivial part is the covariant derivative in the fermion kinetic term, so let us only check that part. 
\bea
U(x) (\diff + \im\, q A)_y U(x)^{-1}&=& (\diff + \im\, q A+2\pi \im\, \diff \Theta(y-x))_y\nonumber\\
&=& e^{ -2\pi \im\, \Theta(y-x) }
\left( \diff + \im\, q A \right)_y  e^{ 2\pi \im\, \Theta(y-x) } . 
%&=& \exp\left(-2\pi \im\, \Theta(y-x)\right) (\diff + \im\, q A)_y \exp\left(2\pi \im\, \Theta(y-x)\right). 
\eea
As $\exp \left( 2\pi\im\, \Theta(y-x) \right) =1$ almost everywhere, we have checked the commutation relation. In the lattice regularized formulation, we can see this more explicitly without any subtlety. 

Because of $\mathbb{Z}_q^{[1]}$ in $(1+1)$d, the Hilbert space decomposes into the $q$ sectors, 
\be 
\mathcal{H}=\bigoplus_{k=0}^{q-1} \mathcal{H}_k, 
\ee
where each Hilbert space $\mathcal{H}_k$ is defined by 
\begin{equation}
    \mathcal{H}_k=\left\{\Psi\in \mathcal{H}\, | \, U(x) \Psi
    = \rme^{\frac{2\pi \im k}{q} } \Psi\right\}. 
\end{equation}
We note that, due to the topological nature of $U(x)$, the condition does not depend on $x$. 
Here, the label $k$ is identified in mod $q$, i.e. $k\sim k+q$. 

We can fix the label $k$ by gauging the $1$-form symmetry. 
Since the $1$-form symmetry is $\mathbb{Z}_q\subset U(1)$, the $2$-form gauge field $\mathcal{B}$ is now subject to the constraint,
\begin{equation}
    q\, \mathcal{B}=\diff \mathcal{C}, 
\end{equation}
where $\mathcal{C}$ is an auxiliary $U(1)$ $1$-form gauge field. 
Now, the $1$-form gauge transformation is given by 
\begin{equation}
    A\mapsto A+\Lambda,\quad 
    \mathcal{B}\mapsto \mathcal{B}+\diff \Lambda,\quad 
    \mathcal{C}\mapsto \mathcal{C}+ q\, \Lambda, 
\end{equation}
and thus the gauge-invariant combinations are given as $F-\mathcal{B}$ and $q\,A-\mathcal{C}$. 
Using this $1$-form gauge transformation, we can set $A=0$ as a gauge-fixing condition, which effectively replace $A$ by ${1\over q}\mathcal{C}$. Then, we obtain the replacement $g^2\to q^2 g^2$ and $\theta_0\to (\theta_0+2\pi k)/q$, where $k$ is the label for the discrete $\theta$-term. 
In this way, we can extract the Hilbert space $\mathcal{H}_k$ by gauging $\mathbb{Z}_q$ including the discrete $\theta$-term. 
That is, the $U(1)$ gauge theory with charge-$q$ matters can be identified as the discrete sum of $U(1)$ gauge theories with charge-$1$ matters with rescaled coupling $q^2 g^2$ and with different fractionalized $\theta$-angles $(\theta_0+2\pi k)/q$. 
Each sector $\mathcal{H}_k$ of the decomposition is recently called as universe~\cite{Tanizaki:2019rbk, Nguyen:2021naa, Komargodski:2020mxz, Cherman:2020cvw}. 

These different universes can be connected by introducing Wilson line operators. 
When a charge-$q_p$ Wilson loop is introduced in the $k$-th universe, its string tension $\sigma$ is given by 
\begin{equation}
    \sigma_{q_p,k}= E_{k-q_p}(\theta_0)-E_k(\theta_0), \label{eq:theo-predict-tension}
\end{equation}
where $E_k(\theta_0)$ describes the ground-state energy density\footnote{
It is the lowest eigenvalue of $H(x)$ and
independent of $x$ for periodic boundary condition because of translational invariance.
} for $\mathcal{H}_k$, $k=1,\ldots, q$.  
When the fermion mass $m^2$ is small enough compared with the photon mass $\mu^2=q^2 g^2/\pi$, 
the ground state energy density can be approximated as 
\begin{equation}
E_k(\theta_0)
=-m{\rme^\gamma q g\over 2\pi^{3/2}} \cos\left({\theta_0-2\pi k\over q}\right)
+\mathcal{O}(m^2 ) . 
\label{eq:q_Schwinger_energy_small}
\end{equation}
%On the other hand, 
When the fermion mass $m$ is large enough, 
the theory is approximately the pure Maxwell theory with theta angle $\theta_0$ and
the ground state energy density behaves as 
\begin{equation}
    E_k(\theta_0)
    ={g^2\over 2}\min_{\ell\in \mathbb{Z}}\left(k+q\ell-{\theta_0\over 2\pi}\right)^2
     +\mathcal{O}\left( m^{-2} \right) . 
    \label{eq:q_Schwinger_energy_large}
\end{equation}
When $q=1$, the Wilson loop does not change the universe, and thus any integer electric charge can be screened by the pair creation. 
When $q>1$, this is not the case, and the Wilson loop obeys an area law for generic values of $\theta_0$ and $q_p\not \in q\mathbb{Z}$. 

Let us point out that the string tension does not have to be positive semi-definite, and the appearance of negative string tensions is suggested by the formula~\eqref{eq:theo-predict-tension} with \eqref{eq:q_Schwinger_energy_small} or \eqref{eq:q_Schwinger_energy_large}. 
However, one might think at the first sight that this is counter-intuitive: 
the string tension in $(1+1)$d is the difference of the energy densities between inside and outside of the Wilson loop, and, since the outside of the loop is the ground state, we necessarily have the positive string tension. 
How can we create states that have lower energy density than ground state?
Here, the notion of the universe~\cite{Tanizaki:2019rbk, Nguyen:2021naa, Komargodski:2020mxz, Cherman:2020cvw} plays an important role. 
Since $U(x)$ is a topological local operator, 
any physical processes cannot change its value as long as the $1$-form symmetry is preserved. 
This means that any processes are closed inside one of the universes. 
Especially, if we have a ground state of a specific universe, 
then it never decays even if it has a larger energy density than another universe. 
Since the Wilson loop connects different universes as we have discussed, 
it is possible to have negative string tensions. 
We also comment on when the string tension changes its sign.
From the equations \eqref{eq:q_Schwinger_energy_small} and \eqref{eq:q_Schwinger_energy_large},
we see that the sign of the string tension changes at the same $(q, q_p ,\theta_0 )$
at least in small and large mass regime.
Therefore it would be natural to expect that
the condition for having the negative string tension is independent of $m$
as long as it is nonzero.

When we take the massless limit of the fermion, $m=0$, 
the formula~\eqref{eq:q_Schwinger_energy_small} suggests the presence of $q$ degenerate vacua and the deconfinement of the Wilson loops. 
In this limit,
the charge-$q$ Schwinger model enjoys the $\mathbb{Z}_q$ chiral symmetry, 
$\psi\mapsto \rme^{{2\pi\over 2q}\im \gamma_3}\psi$ 
and $\overline{\psi}\mapsto \overline{\psi}\,\rme^{{2\pi\over 2q}\im \gamma_3}$. 
This transformation may look to be $\mathbb{Z}_{2q}$ instead of $\mathbb{Z}_q$, 
but we note that the fermion parity is a part of the $U(1)$ gauge redundancy, 
so the proper symmetry transformation is just $\mathbb{Z}_q$. 
The vacuum degeneracy is the consequence of its spontaneous breaking, 
and it is required to match the 't~Hooft anomaly between $\mathbb{Z}_q^{[1]}$ 
and $\mathbb{Z}_q$ chiral symmetry 
with nonzero mass gap~\cite{Anber:2018jdf, Anber:2018xek, Armoni:2018bga, Misumi:2019dwq}. 

Even with nonzero $m$, 
we have an 't~Hooft anomaly, and/or global inconsistency~\cite{Gaiotto:2017yup, Tanizaki:2017bam, Kikuchi:2017pcp, Tanizaki:2018xto, Karasik:2019bxn, Cordova:2019jnf, Cordova:2019uob}, 
to constrain the $\theta_0$ dependence of low-energy physics of the charge-$q$ Schwinger model~\cite{Misumi:2019dwq}. 
As we can see explicitly in the formula~\eqref{eq:q_Schwinger_energy_small} of the ground state energy, 
each ground state $E_k(\theta_0)$ does not have the $2\pi$ periodicity of $\theta_0$. 
Instead, it satisfies $E_k(\theta_0+2\pi)=E_{k-1}(\theta_0)$, 
and thus the energy spectrum has the $2\pi$ periodicity thanks to the multi-branch structure of the ground states. 
This nicely imitates the situation of the $(3+1)$d pure Yang-Mills theory~\cite{Witten:1980sp,tHooft:1981bkw, Witten:1998uka}.\footnote{
However, there is an important difference between the $2$d charge-$q$ Schwinger model and $4$d $SU(N)$ pure Yang-Mills (YM) theory regarding the selection rule. 
As we have emphasized above, charge-$q$ Schwinger model has $\mathbb{Z}_q^{[1]}$, and its Hilbert space on $S^1$ is decomposed into $q$ universes. 
Each branch of the $q$ vacua belong to different universes, so they are stable. 
On the other hand, 
the $4$d YM theory also has $\mathbb{Z}_N^{[1]}$, but $4$d QFTs with $1$-form symmetries do not have such decomposition. 
Therefore, only the true ground state is stable, and the other $(N-1)$ vacua experience the vacuum decay by creation of domain walls. 
} 
This level crossing of the ground states is required to satisfy the anomaly 
and/or global inconsistency matching conditions. 

%%%%%%%%%%%%%%%%
%%%%%%%%%%%%%%%%
\subsection{Charge-$q$ Schwinger model on $[0,L]$ and $\theta$-periodicity}\label{sec:periodicity}
%%%%%%%%%%%%%%%%
%%%%%%%%%%%%%%%%

Now, let us discuss what happens if we consider the theory on the interval, $[0,L]$. 
As discussed in the following sections, 
our strategy of quantum computation for Schwinger model is designed for the open boundary condition, not for the periodic boundary condition. 
We need to understand how the physics changes by the choice of the boundary condition.\footnote{
In Appendix~\ref{sec:Maxwell_canonical_interval},  
this problem is discussed for a simpler setup, the $(1+1)$d pure Maxwell theory. 
Many of the following discussion is made more explicit thanks to the exact solvability of the model. }

To discuss gauge theories with the boundaries, we have to specify the charges at the boundaries. 
When the external charges $\pm k\in \mathbb{Z}$ are put on the boundaries, 
the Gauss law constraint is given by 
\begin{equation}
    \p_1 \Pi(x)+k\left( \delta(x-L)-\delta(x)\right) 
    = q\, \psi^\dagger \psi(x). 
\end{equation}
Let us denote the corresponding Hilbert space as $\mathcal{H}_k^{(\mathrm{open})}$. 
The Gauss law above tells us that
the $1$-form symmetry generator $U(x)=\exp\left({2\pi\im\over q}\Pi(x)\right)$ 
is now completely fixed upon acting on the physical states
and its value is given by 
\begin{equation}
    U(x)= \exp\left({2\pi\im\over q}k\right), 
    \label{eq:interval_sector_choice}
\end{equation}
as we do not introduce any test charges except at the boundaries. 
This point is drastically different from the case with the periodic boundary condition. 
In the periodic boundary condition, $U(x)$ can take $q$ distinct values, and the Hilbert space is decomposed into $q$ universes by the eigenvalue of $U(x)$. 
In some sense, on the open interval, the projection to the single universe is automatically selected by the choice of the boundary condition. 

Let us discuss how these properties affect the $\theta$-angle periodicity, $\theta_0\sim \theta_0+2\pi$. 
For the periodic boundary condition, we can use the spatial Wilson loop operator, $W(S^1)=\exp\left(\im \oint A_1(x)\diff x\right)$, as the gauge-invariant unitary operator, and we can relate $\theta_0+2\pi$ and $\theta_0$ by the unitary transformation of $W(S^1)$. 
This ensures the $2\pi$ periodicity of the spectrum for any $U(1)$ gauge theories with the periodic boundary condition, and the formula~\eqref{eq:q_Schwinger_energy_small} gives an explicit realization. 
For the open boundary condition, however, we do not have such a gauge-invariant unitary operator. 
Therefore, nothing ensures that physical quantities have $2\pi$ periodicity with respect to $\theta_0$.  
Indeed, we can see that $2\pi$ periodicity is completely lost, $\theta_0\not\sim \theta_0+2\pi$, when we consider the charge-$q$ Schwinger model with the open boundary. 

For the charge-$q$ Schwinger model, the $\theta$-angle periodicity can be found for a single sector $\mathcal{H}_k$ for bulk local observables, and that periodicity is given by $\theta_0\sim \theta_0+2\pi q$. 
This is a natural expectation based on the fact that $1$-form symmetry group is just $\mathbb{Z}_q$ and we do not have other $1$-form symmetries. 
For example, let us find the ground-state energy density, 
\begin{equation}
    E_k^{(\mathrm{open})}(x,\theta_0)=\langle\mathrm{GS},k| H(x)|\mathrm{GS},k\rangle. 
\end{equation}
Since $U(x)|\mathrm{GS},k\rangle=\rme^{2\pi \im k/q}|\mathrm{GS},k\rangle$, $|\mathrm{GS},k\rangle$ should behave as the ground state of the $k$-th universe if the volume is sufficiently large. 
When the location $x$ is sufficiently far from the both ends, i.e., $0\ll x\ll L$, 
it is given by the same formula (\ref{eq:q_Schwinger_energy_small}) or (\ref{eq:q_Schwinger_energy_large}) 
up to exponentially small corrections:
\begin{eqnarray}
E_k^{(\mathrm{open})}(x,\theta_0)=E_k(\theta_0)+\mathcal{O}(\rme^{-\mu|x|}, \rme^{-\mu(L-|x|)}), 
\label{eq:energy_density_local}
\end{eqnarray}
where $\mu$ denotes the mass gap of the theory, which is given by $\mu=qg/\sqrt{\pi}$ 
in the massless fermion limit, $m\to 0$. 
The exponentially small terms represent the boundary effects and need not have any $\theta_0$'s periodicity. 
Only when we can neglect the boundary contributions, we can observe the manifest $2\pi q$ periodicity, $E_k(\theta_0+2\pi q)=E_k(\theta_0)$. 

Here, we would like to emphasize the importance of using the local density $H(x)$. 
Another popular way to compute the energy density is that we calculate the total energy first and divide it by the volume, however the result becomes exponentially worse. 
In order to see this, let us compute the total energy 
\begin{align}
    E_k^{\mathrm{total}}(\theta_0)
    =\int_0^L \diff x\, E^{(\mathrm{open})}(x,\theta_0)
    & =\int_0^L\diff x\left(E_k(\theta_0)+\mathcal{O}(\rme^{-\mu|x|}, \rme^{-\mu(L-|x|)})\right)\nonumber\\
    &=LE_k(\theta_0)+\mathcal{O}(1),
\end{align}
where the second term on the right hand side represents the localized energy around the boundary. 
If we try to obtain the energy density as 
\begin{equation}
    {1\over L}E_k^{\mathrm{total}}(\theta_0)=E_k(\theta_0)+\mathcal{O}(L^{-1}), 
\end{equation}
then we really have to take the infinite volume limit to suppress the boundary effect, $\mathcal{O}(L^{-1})$. 
However, since the system is generically gapped including the massless point, $m=0$, we should be able to calculate $E_k(\theta_0)$ using the open boundary condition with exponentially good accuracy, and \eqref{eq:energy_density_local} realizes this physical intuition. 
The same comment applies also for other quantities, such as chiral condensates.

%%%%%%%%%%%%%%%%
%%%%%%%%%%%%%%%%
%%%%%%%%%%%%%%%%
\section{Lattice formulation of charge-$q$ Schwinger model}
\label{sec:lattice}
%%%%%%%%%%%%%%%%
%%%%%%%%%%%%%%%%
%%%%%%%%%%%%%%%%
Now, let us consider a lattice formulation of the charge-$q$ Schwinger model on finite interval $[0,L]$. 
Here, we rewrite the Hamiltonian \eqref{eq:hamiltonian-cont} in terms of the spin operators acting on the qubits.
In this and next sections, we basically follow  the notation and strategy in Ref.~\cite{Honda:2021aum}, where the Schwinger model with $q=1$ is discussed.

%%%%%%%%%%%%%%%%
%%%%%%%%%%%%%%%%
\subsection{Charge-$q$ Schwinger model using staggered fermion}
%%%%%%%%%%%%%%%%
%%%%%%%%%%%%%%%%
Let us put the theory on a lattice with $N$ sites and lattice spacing $a$.
We realize the two-component Dirac fermion $\psi (x)$ using the staggered fermion $\chi_n$,
where $n$ labels the lattice site ($x=na$). 
Here, $\chi_n$ is a single component complex fermionic operator, and the Dirac fermion at $x$ extends over two sites on the lattice:
\begin{equation}
{1\over \sqrt{a}} 
\begin{pmatrix}    
\chi_{2\lfloor n/2\rfloor} \cr \chi_{2\lfloor n/2\rfloor+1}
\end{pmatrix}
\leftrightarrow \psi(x). 
\end{equation}
The gauge field and its canonical momentum are represented by the link variables,
\begin{\eq}
U_n \leftrightarrow e^{\im a  A_1 (x)} ,\quad L_n \leftrightarrow -\Pi(x),		
\end{\eq}
which are defined on the link between the sites $n$ and $n+1$.
We will take open boundary conditions for the fields.

The lattice discretization of the Hamiltonian~\eqref{eq:hamiltonian-cont} is given by
\begin{\eqa}
H 
= J \sum_{n=0}^{N-2} \left( L_n +\frac{\theta_0}{2\pi}  \right)^2 
 -\im w \sum_{n=0}^{N-2} \Bigl[ \chi_n^\dag (U_n)^q \chi_{n+1} -{\rm h.c.} \Bigr]  +m\sum_{n=0}^{N-1} (-1)^n \chi_n^\dag \chi_n ,
\end{\eqa}
where the parameters $J$ and $w$ are defined as
\begin{\eq}
J=\frac{g^2 a}{2} ,\quad  w=\frac{1}{2a} .
\end{\eq}
The $\theta$-term is realized by a constant shift of link variable $L_n$, so that it can be interpreted as an constant electric flux.
No additional difficulty appears to perform the numerical simulation in non-zero $\theta_0$ regime
in contrast to the conventional Monte Carlo approach.\footnote{
In the usual Wilson lattice formulation of the Euclidean path integral, 
the topological $\theta$ term gives an imaginary phase to the Boltzmann weight, 
and we suffer from the sign problem. 
For $(1+1)$d $U(1)$ gauge theories, 
this problem is recently resolved by the combination of Villain lattice formulation and the worm algorithm for the worldline method~\cite{Gattringer:2018dlw}. 
}
The field operators obey the canonical commutation relations:
\begin{\eq}
[L_n, U_m] = U_m\delta_{nm} ,\quad \{ \chi_n ,\chi_m^\dag \} = \delta_{nm}.
\end{\eq}

On a lattice, the Gauss law (\ref{eq:Gauss_continuum}) takes the form,
\begin{\eq}
L_n -L_{n-1} 
= q \Biggl[ \chi_n^\dag \chi_n -\frac{1-(-1)^n}{2}  \Biggr].
\end{\eq}
With the open boundary condition, we can solve it as
\begin{\eq}
L_n = L_{-1} +q \sum_{j =1}^n \left( \chi_j^\dag \chi_j -\frac{1-(-1)^j }{2} \right) .
\end{\eq}
Note that the $q$ dependence appears in the Gauss law. 
We impose a boundary condition on the electric field operator by setting $L_{-1}=0$, which corresponds to the choice $k=0$ in (\ref{eq:interval_sector_choice}).
Upon fixing the gauge so that $U_n=1$ for all $n$, we have the Hamiltonian as follows,
\begin{\eq}
H 
= -\im w \sum_{n=1}^{N-1} \Bigl[ \chi_n^\dag  \chi_{n+1} -{\rm h.c.} \Bigr] 
 +m\sum_{n=1}^N (-1)^n \chi_n^\dag \chi_n  
 +J \sum_{n=1}^N \Biggl[\frac{\theta_0}{2\pi} +q \sum_{j =1}^n \left( \chi_j^\dag \chi_j -\frac{1-(-1)^j}{2} \right)  \Biggr]^2 .
\label{eq:lattice_hamiltonian_0}
\end{\eq}
Notice that this Hamiltonian only contains the fermionic operators 
and thus the local Hilbert space is finite dimensional. 

We note that the $2\pi$ periodicity of $\theta_0$ is lost by taking the open boundary condition. 
If we take the periodic boundary condition instead, then we can define the gauge-invariant unitary operator, $W=\prod_n U_n$, which is nothing but the spatial Wilson loop, and we have $H|_{\theta_0+2\pi}=W^{-1} (H|_{\theta_0})W$. 
For the periodic boundary condition, however, we necessarily have an infinite-dimensional local Hilbert space, 
and we cannot map to the system only with fermions\footnote{
For such a case, we need to make further a truncation on the Hilbert space.
}.  
Since we would like to realize this system on the quantum computer, locally finite Hilbert space is more natural, so we take the open boundary condition in the following.

%%%%%%%%%%%%%%%%
\subsection{Insertion of the probes}
\begin{figure}[t]
\centering
\includegraphics[scale=0.45]{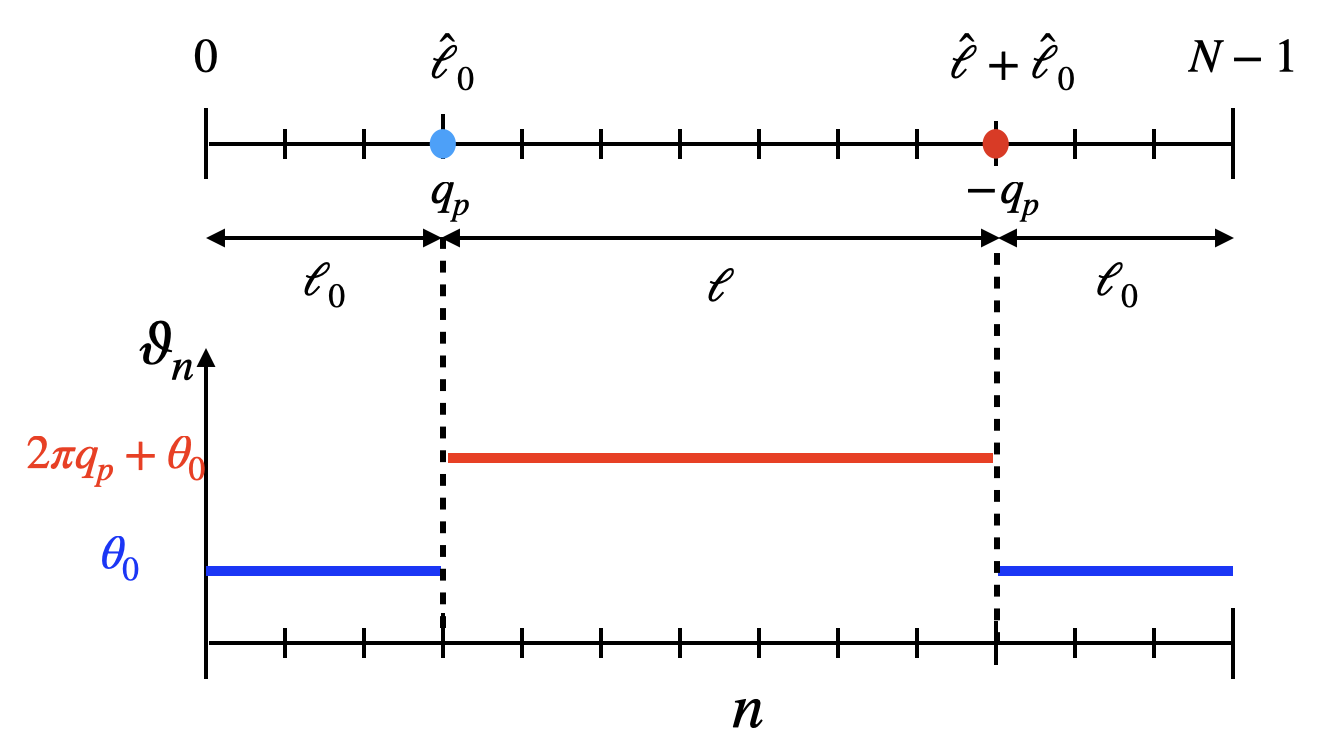}
\caption{%
Correspondence of the insertion of probe charge and Wilson loop. 
}
\label{fig:Image-Wloop}
\end{figure}

%%%%%%%%%%%%%%%%
We introduce the two probe charges $+q_p$ and  $-q_p$ on the $\hat{\ell}_0$-th and $(\hat{\ell}_0+\hat{\ell})$-th sites, respectively. Here, $\hat{\ell}$ $(\hat{\ell}_0)$ denotes the dimensionless quantity, $\hat{\ell} \equiv \ell/a$ $(\hat{\ell}_0 \equiv \ell_0/a)$.
This gives a constant shift for the link variable between $\hat{\ell}_0$-th and $(\hat{\ell}_0+\hat{\ell})$-th sites, which corresponds to the insertion of long rectangular Wilson loop with width $\ell$ as shown Fig.~\ref{fig:Image-Wloop}.
%%%%%%%%%%%%%%%%
%%%%%%%%%%%%%%%%
It is realized by introducing the position-dependent $\theta$-angle,
\begin{align}
\label{eq:def-vartheta}
\vartheta_n 
= \left\{
 \begin{array}{cc}
  \theta_0 +2\pi q_p  & \ {\rm for}\ \hat{\ell}_0 \le n < \hat{\ell}_0 +\hat{\ell} \\
  \theta_0 & \text{otherwise}
 \end{array} .
 \right.
\end{align}
For open boundary conditions, it would be appropriate to take
\begin{\eq}
\hat{\ell}_0 = \frac{N-\hat{\ell} -1}{2}  ,
\end{\eq}
with odd $\hat{\ell}$ for even $N$ and even $\hat{\ell}$ for odd $N$. 
In this way, we equally separate the probe charges from both boundaries, so that the boundary effect is suppressed as much as possible in the simulation with finite volumes.

In the presence of the probes, the lattice Hamiltonian is
\begin{\eq}
H 
= -\im w \sum_{n=1}^{N-1} \Bigl[ \chi_n^\dag  \chi_{n+1} -{\rm h.c.} \Bigr] 
 +m\sum_{n=1}^N (-1)^n \chi_n^\dag \chi_n  
 +J \sum_{n=1}^N \Biggl[\frac{\vartheta_n }{2\pi} +q \sum_{j =1}^n \left( \chi_j^\dag \chi_j -\frac{1-(-1)^j}{2} \right)  \Biggr]^2 ,
\end{\eq}
where $\theta_0$ in \eqref{eq:lattice_hamiltonian_0} is replaced by the position dependent one, $\vartheta_n$ in \eqref{eq:def-vartheta}. 
Using the Jordan-Wigner transformation,
\begin{align}
\chi_n =  \frac{X_n -\im Y_n}{2} \left( \prod_{i=1}^{n-1} -\im Z_i \right),~~
\chi_n^\dag =  \frac{X_n +\im Y_n}{2} \left( \prod_{i=1}^{n-1} \im Z_i \right),
\end{align}
we obtain the Hamiltonian in terms of spin operators,
\begin{align}
 H = J\sum_{n=0}^{N-2} \left[ q \sum_{i=0}^{n}\frac{Z_i + (-1)^i}{2}+\frac{\vartheta_n}{2\pi}\right]^2 
 + \frac{w}{2}\sum_{n=0}^{N-2}\big[X_n X_{n+1}+Y_{n}Y_{n+1}\big]
 + \frac{m}{2}\sum_{n=0}^{N-1}(-1)^n Z_n ,
\end{align}
where ($X_n,Y_n,Z_n$) stands for the Pauli matrices ($\sigma^1, \sigma^2,\sigma^3$) at site $n$.
Here, we have dropped irrelevant constants independent of $\vartheta_n$ in the Hamiltonian.
The form of the spin Hamiltonian indicates the convenient relation,
\begin{\eq}
H \left( q , J ,  \vartheta_n \right)
= H\left( 1 , q^2 J ,  \frac{\vartheta_n}{q} \right) ,
\label{eq:relation-q-Schwinger}
\end{\eq}
which enables us to relate the $q=1$ case to general $q$ case.
We would like to emphasize that this translation to $q=1$ is possible because we take the open boundary condition. 
When we take the periodic boundary condition, we cannot eliminate the spatial link variables completely by gauge fixing. 
Then, the spatial hopping term, $\chi^\dagger_n(U_n)^q\chi_{n+1}$, genuinely depends on the choice of $q\ge 1$, and we cannot relate them by simple replacements of coupling constants.
As we have discussed in Sec.~\ref{sec:chargeq_Schwinger_1form}, results for charge $q$ can be technically translated in the language of the $q =1$ case
with fractional probe charge, but it necessarily accompanies the projection to the single universe. 
It requires extra modification of the large gauge invariance to keep the global information of the theory with periodic boundary condition.

%%%%%%%%%%%%%%%%
%%%%%%%%%%%%%%%%
%%%%%%%%%%%%%%%%
\section{Simulation strategy}
\label{sec:strategy}
%%%%%%%%%%%%%%%%
%%%%%%%%%%%%%%%%
%%%%%%%%%%%%%%%%
We would like to study the expectation values of physical operators $\mathcal{O}$,
\beq
  \langle \mathcal{O}  \rangle= \langle {\rm GS}| \mathcal{O} | {\rm GS} \rangle, 
\eeq
where $|{\rm GS}\rangle$ denotes the ground state of the full Hamiltonian.
In this section, we will first explain how to obtain $|{\rm GS}\rangle$ using the adiabatic state preparation and also show the outline to design its quantum circuit.
After that we discuss the way to measure observables 
such as the total energy and local energy density
in the framework of digital quantum simulation.

%%%%%%%%%%%%%%%%
%%%%%%%%%%%%%%%%
\subsection{Adiabatic state preparation of vacuum}
%%%%%%%%%%%%%%%%
%%%%%%%%%%%%%%%%
In the adiabatic state preparation, the first step is to choose an initial Hamiltonian $H_0$ 
whose ground state $| {\rm GS}_0 \rangle$ is unique and known.
The second step is to choose a time-dependent adiabatic Hamiltonian $H_A (t)$ such that
\begin{align}
H_\mathrm{A} (0) = H_0, \quad H_\mathrm{A} (T)=H.
\end{align}
Then the adiabatic theorem claims that we can construct the ground state
of the target Hamiltonian $H$:
if the system with the Hamiltonian $H_A (t)$ has a unique gapped ground state for any $t\in[0,T]$, 
then the ground state $|{\rm GS} \rangle$  is obtained by~\cite{messiah1962quantum,farhi2000quantum}
\begin{align}
|{\rm GS}  \rangle 
= \lim_{T\rightarrow\infty} \mathcal{T}\exp{\left(-\im\int_0^T \diff t\ H_\mathrm{A} (t) \right)} 
|{\rm GS}_0  \rangle .\label{eq:adiabatic-prep}
\end{align}
Here, the symbol $\mathcal{T}$ indicates the time ordering of subsequent operators.

In the present study, we choose the initial Hamiltonian as
\begin{\eq}
H_0=\left.H\right|_{J=w=\vartheta_n=0,\, m=m_0},
\end{\eq} 
for some $m_0>0$.
The ground state of $H_0$ is given by the N\'eel ordered state,
\begin{align}
|\rm{GS}_0 \rangle =\ket{1010 \dots}
:=\ket{1}\otimes\ket{0}\otimes\ket{1}\otimes\ket{0}\otimes\cdots
\end{align}
with $Z|0 \rangle =+|0 \rangle$ and $Z|1 \rangle = -|1 \rangle$,
and it can be easily constructed by
\begin{align}
|{\rm GS}_0 \rangle 
=\prod_{j=0}^{\lfloor
\frac{N-1}{2} \rfloor
} X_{2j} |00
\dots \rangle .
\end{align}
Here, $\lfloor x\rfloor$ is the floor function of $x$, which denotes the largest integer no greater than $x$.

We express the unitary evolution in \eqref{eq:adiabatic-prep} using the Suzuki-Trotter product formula. 
Considering the following decomposition of the Hamiltonian turns out to be useful~\cite{Honda:2021aum}:
\begin{\eq}
 H = H_{XY}^{(0)} + H_{XY}^{(1)} + H_{Z} + C,\label{eq:H_decomposition0}
\end{\eq}
where, for odd $N$,
each term is
\begin{align}
\label{eq:H_decomposition-each}
\begin{split}
 &H_{XY}^{(0)} = \frac{w}{2} \sum_{n=0}^{\frac{N-3}{2}}\big(X_{2n}X_{2n+1}+Y_{2n}Y_{2n+1}\big),\\
& H_{XY}^{(1)} = \frac{w}{2} \sum_{n=1}^{\frac{N-1}{2}}\big(X_{2n-1}X_{2n}+Y_{2n-1}Y_{2n}\big),
 \\
  &H_{Z} 
  = \frac{q^2 J}{2}\sum_{n=0}^{N-3} \sum_{k=n+1}^{N-2}(N-k-1)Z_n Z_k
  +\frac{q^2 J}{2}\sum_{n=0}^{N-2}
  \frac{1+(-1)^n}{2}  \sum_{i=0}^{n}Z_i
 +q_p q J \sum_{k=0}^{\hat{\ell}_0+\hat{\ell}-1} (\hat{\ell}_0+\hat{\ell}-k)Z_k  \\
&
\qquad\qquad
-q_p q J \sum_{k=0}^{\hat{\ell}_0-1}(\hat{\ell}_0-k)Z_k
+\frac{q \theta_0}{2\pi}J\sum_{k=0}^{N-2}(N-k-1)Z_k+ \frac{m}{2}\sum_{n=0}^{N-1}(-1)^n Z_n,
\\
&C =
\frac{q_p q J}{2}\left(\hat{\ell}+(-1)^{\hat{\ell}_0}\frac{1-(-1)^{\hat{\ell}}}{2}\right)
+q_p \left( q_p +\frac{\theta_0}{\pi}\right)J\hat{\ell}
+\frac{q \theta_0J}{4\pi}\left(N-1+\frac{1 +(-1)^N}{2}\right)\\
& \qquad\qquad + \left(\frac{\theta_0}{2\pi}\right)^2J(N-1).
\end{split}
\end{align}
Here, $C$ is just a constant, but it is important to compute the $\theta_0$ dependence of the ground state energy. 
The summands of $H_{XY}^{(0)}$ commute with one another, 
and thus we can obtain the unitary operator, 
$\exp\left( -\im \varepsilon H_{XY}^{(0)} \right)$, 
by taking the product of local unitary operations, 
$\exp\left( -\im \varepsilon {w\over 2}(X_{2n}X_{2n+1}+Y_{2n}Y_{2n+1})\right)$, without care about ordering. 
The same is true for $H_{XY}^{(1)}$ and $H_Z$. 

Combining the adiabatic theorem and the product formula, we find the approximate form of the ground state:
\begin{align}
 |{\rm GS_A} \rangle \
  :=
 & \
\prod_{r=1}^{M} \Bigl(
 e^{-\im H_{XY,r}^{(0)}\frac{\delta t}{2}} e^{-\im H_{XY,r}^{(1)}\frac{\delta t}{2}} e^{-\im H_{Z,r}\delta t}
 e^{-\im H_{XY,r}^{(1)}\frac{\delta t}{2}} e^{-\im H_{XY,r}^{(0)}\frac{\delta t}{2}}
 \Bigr)
|{\rm GS}_0 \rangle. \
\label{eq:approximate_GS}
\end{align}
Here, $M:=T/\delta t$ is a positive integer, which should be taken to be large in practice for good approximation.
This is the second-order Suzuki-Trotter approximation, and its error is estimated as $\mathcal{O}(M\delta t^2)$ for the whole operator for fixed $T$~\cite{Lloyd1073,Suzuki1991}.
Here, $H_{XY,r}^{(0)}$, $H_{XY,r}^{(1)}$, and $H_{Z,r}$ are obtained by the following replacements in \eqref{eq:H_decomposition-each},
\begin{align}
& w\to w f\left( \frac{r}{M} \right), \quad
 \theta_0\to \theta_0 f\left( \frac{r}{M} \right), \nonumber\\
& q_p \to q_p f\left( \frac{r}{M} \right),   \quad
 m\to m_0 \left( 1-f\left( \frac{r}{M} \right) \right) + mf\left( \frac{r}{M} \right),
\label{eq:ad-fn}
\end{align}
where the function $f(s)$ called adiabatic schedule is smooth and satisfies
\begin{\eq}
f(0)=0 ,\quad f(1)=1 .
\end{\eq}
There are infinitely many choices of $f(s)$ and
any choice of $f(s)$ works in principle 
as long as the setup satisfies the assumptions of the adiabatic theorem and
$T$ is sufficiently large.
However, choice of $f(s)$ affects accuracy of the approximation in practice. 
In Appendix~\ref{sec:Ad-fn},
we study various choices of adiabatic schedules at some values of the parameters and
identify the best one among them.
As a conclusion, we choose the adiabatic schedule as
\begin{\eq}
f(s)=\frac{\tanh (s)}{\tanh (1)} .
\end{\eq}

%%%%%%%%%%%%%%%%
%%%%%%%%%%%%%%%%
\subsection{Quantum simulation protocol for Hamiltonian evolution}
%%%%%%%%%%%%%%%%
%%%%%%%%%%%%%%%%
In this subsection, let us quickly mention the quantum circuits to obtain $|\mathrm{GS}_\mathrm{A}\rangle$. 
The universality theorem tells that
one can approximate any unitary operations on multiple qubits by combination of simple qubit gates and the controlled NOT (CNOT) gate. 
Indeed, our expression~\eqref{eq:approximate_GS} only contains two-quibit unitary transformations, so we just have to express them.  
Here, we use the following three operations as elementary gates to realize the unitary evolution~\eqref{eq:approximate_GS}, 
\begin{itemize}
\item Hadamard gate
\begin{align}
\label{eq:gate_H}
&\begin{array}{c}
\Qcircuit @C=0.5cm @R=.3cm {
&  \gate{H} & \qw
}
\end{array}=\frac{1}{\sqrt{2}}
\begin{pmatrix}
 1 & 1 \\ 1 & -1
\end{pmatrix},
\end{align}
\item $Z$-rotation gate
\begin{align}
\label{eq:gate_Zrot}
&\begin{array}{c}
\Qcircuit @C=0.5cm @R=.3cm {
&  \gate{R_Z(\theta)} & \qw
}
\end{array}
= \e^{-\im\frac{\theta}{2}Z}=
\begin{pmatrix}
 \e^{-\im\frac{\theta}{2}} & 0 \\ 0 & \e^{\im\frac{\theta}{2}}
\end{pmatrix},
\end{align}
\item CNOT (CX) gate
\begin{align}
\label{eq:gate_CNOT}
\begin{array}{c}
\Qcircuit @C=1em @R=.7em {
& \ctrl{1} & \qw
\\
& \targ{} & \qw
}
\end{array}
= |0\rangle \langle 0| \otimes I+|1\rangle \langle 1|\otimes X
=\begin{pmatrix} 
1 & 0  & 0 & 0 \cr  
0 & 1  & 0 & 0 \cr  
0 & 0  & 0 & 1 \cr  
0 & 0  & 1 & 0 \cr  
\end{pmatrix},
\end{align}
where $I$ stands for the identity operator.
\end{itemize}
For notational convenience, we also introduce a circuit diagram of the $S$-gate, which is defined by,
\begin{align}
&\begin{array}{c}
\Qcircuit @C=0.5cm @R=.3cm {
&  \gate{S} & \qw
}
\end{array}=
\begin{pmatrix}
 1 & 0 \\ 0 & \im
\end{pmatrix}
=\Qcircuit @C=0.5cm @R=.3cm {
&  \gate{R_Z(\pi/2)} & \qw
}.
\end{align}
The last equality holds up to an overall phase, which does not affect the following discussion.

When trying to generate the ground state by \eqref{eq:approximate_GS}, the only nontrivial step is to realize the two-qubit operations, $\e^{-\im\alpha (X_n X_{n+1} + Y_n Y_{n+1})}$ and $\e^{-\im\alpha Z_n Z_{n+1}}$, using the above elementary gates. 
These operations can be expressed as 
\begin{align}
\label{eq:XY_circuit}
\e^{-\im\alpha (X_n X_{n+1} + Y_n Y_{n+1})}
&= 
 %\hspace{3em}
\begin{array}{c}
\Qcircuit @C=0.2cm @R=.3cm {
%\lstick{n\text{{=0}}}
&
\gate{H}&\gate{S}&\ctrl{1} & \gate{H} &\gate{R_Z(2\alpha)} & \gate{H} &\ctrl{1} &\gate{S^\dag}&\gate{H}&\qw
\\
%\lstick{n\text{=1}}
&
\gate{H}&\gate{S}& \targ &\qw & \gate{R_Z(2\alpha)} &\qw &\targ & \gate{S^\dag} &\gate{H} &\qw
}
\end{array}
%\hspace{-10mm}
\\[1em]
%&
\e^{-\im\alpha Z_{n} Z_{n+1}}
%\nonumber \\
&=
%& 
 %\hspace{3em}
\begin{array}{c}
\Qcircuit @C=0.5cm @R=.3cm {
%\lstick{n\text{=0}}
&\ctrl{1} &\qw&\ctrl{1} & \qw
\\
%\lstick{n\text{=1}}
& \targ &  \gate{R_Z(2\alpha)} & \targ & \qw
}
\end{array}
\label{eq:Z_circuit}
\end{align}
Here the top and bottom lines correspond to the site $n$ and $n+1$, respectively, and $\alpha$ is a real parameter.
%

%%%%%%%%%%%%%%%%
%%%%%%%%%%%%%%%%
\subsection{Measurements of physical observables}
%%%%%%%%%%%%%%%%
%%%%%%%%%%%%%%%%
%%%%%%%%%%%%%%%%
\subsubsection{Measurements of total energy} 
\label{sec:measure_H}
%%%%%%%%%%%%%%%%%%%%%%%%%
We discuss how to extract the vacuum expectation value of Hamiltonian and its statistical uncertainty by closely following Appendix B.2 of~\cite{Honda:2021aum} for the sake of self-containedness.
We note that, as discussed in Sec.~\ref{sec:periodicity}, 
the total energy is contaminated from the boundary effect, and it is less useful for the discussion of bulk properties. 
Still, this quantity is useful to compute the inter-particle potential of the probe charges. 
In the next subsection, we shall discuss the way to measure the local densities.

We aim to compute the quantity
\begin{\eq}
 E^{\mathrm{total}}(q_p, \ell,\theta_0 ) = \langle {\rm GS}_A |H (q_p, \ell, \theta_0)|{\rm GS}_A \rangle, 
\end{\eq} 
which approximates the ground state energy.
This involves three independent measurements to compute the expectation values of $H_{XY}^{(0)}$, $H_{XY}^{(1)}$ and $ H_{Z}$ in \eqref{eq:H_decomposition0}. 
Each of them measures a set of operators within which any operator commute to each other.
We execute the circuit for each measurement $n_\text{shots}$ times, 
which leads to the statistical uncertainties in the simulation.
The potential between the probe charges is obtained as 
\beq
V(q_p, \ell, \theta_0) = E^{\mathrm{total}}(q_p, \ell, \theta_0) - E^{\mathrm{total}}(0, 0, \theta_0).
\eeq

We spell out the measurement protocol used to compute the Hamiltonian expectation value and its statistical uncertainty given a desired quantum state has been prepared.  
We consider the case where the state is a 5-qubit state. 
The corresponding lattice sites are labeled by $n\in\{0,1,2,3,4\}$.

The term $H_{XY}^{(0)}$ consists of the operators
\begin{align}
\{X_0X_1, Y_0Y_1, X_2X_3, Y_2Y_3\}.
\end{align}
These operators can be simultaneously measured by noting that
\begin{align}
\begin{split}
 &X_iX_j = CX_{ij}H_i (Z_i I_j) H_i \, CX_{ij},
 \\
 &Y_iY_j = -CX_{ij}H_i (Z_i Z_j) H_i \, CX_{ij}.
\end{split}
\end{align}
The measurement is done with the following circuit:
\begin{align}
\Qcircuit @C=0.5cm @R=.3cm {
\lstick{n\text{=0}}&\ctrl{1} &\gate{H}&\meter
\\
\lstick{n\text{=1}}& \targ &  \qw & \meter
\\
\lstick{n\text{=2}}&\ctrl{1} &\gate{H}&\meter
\\
\lstick{n\text{=3}}& \targ &  \qw & \meter
\\
\lstick{n\text{=4}}& \qw &  \qw & \qw
}
 \nonumber
\end{align}
The input state is supposed to be the state of our interest. 
The four operations at the right end are the classical measurements on the $Z$ basis.
Having obtained the count of each bit string `$b_0b_1b_2b_3$' with $b_i\in\{0,1\}$ from the measurements, 
we calculate the expectation values as 
\begin{align}
\begin{split}
 &\left\langle X_0X_1 \right\rangle = \sum_{b_0,b_1,b_2,b_3}
 (1-2b_0)\frac{\text{counts}_{b_0b_1b_2b_3}}{n_\text{shots}},
 \\
 &\left\langle Y_0Y_1 \right\rangle = -\sum_{b_0,b_1,b_2,b_3}
 (1-2b_0)(1-2b_1)\frac{\text{counts}_{b_0b_1b_2b_3}}{n_\text{shots}},
 \\
  &\left\langle X_2X_3 \right\rangle = \sum_{b_0,b_1,b_2,b_3}
  (1-2b_2)\frac{\text{counts}_{b_0b_1b_2b_3}}{n_\text{shots}},
 \\
 &\left\langle Y_2Y_3 \right\rangle = -\sum_{b_0,b_1,b_2,b_3}
 (1-2b_2)(1-2b_3)\frac{\text{counts}_{b_0b_1b_2b_3}}{n_\text{shots}}.
\end{split}
\nonumber
\end{align}
Here, counts$_{b_0b_1b_2b_3}$ denotes the number of times that the bit string `${b_0b_1b_2b_3}$' is observed.
Defining
\begin{align}
    h_{XY}^{(0)}(b_0,b_1,b_2,b_3) 
    :=\frac{w}{2}\sum_{n=0,2} \Bigl[ (1-2b_n)-(1-2b_n)(1-2b_{n+1}) \Bigr]
\end{align}
for each bit string, we have the expectation value of $H_{XY}^{(0)}$ in the concise form,
\begin{align}
    \left\langle H_{XY}^{(0)} \right\rangle
    =\sum_{b_0,b_1,b_2,b_3}h_{XY}^{(0)}(b_0,b_1,b_2,b_3) 
    \frac{\text{counts}_{b_0b_1b_2b_3}}{n_\text{shots}}.
\end{align}
The same numerical data also allows us to compute 
the expectation value of $\left( H_{XY}^{(0)} \right)^2$,
\begin{align}
    \left\langle \left( H_{XY}^{(0)} \right)^2 \right\rangle
    =\sum_{b_0,b_1,b_2,b_3}(h_{XY}^{(0)}(b_0,b_1,b_2,b_3))^2
    \frac{\text{counts}_{b_0b_1b_2b_3}}{n_\text{shots}},
\end{align}
which is used to estimate the statistical uncertainty.

The term $H_{XY}^{(1)}$ consists of the operators 
\begin{align}
\left\{ X_1X_2, Y_1Y_2, X_3X_4, Y_3Y_4 \right\}.
\end{align}
Hence, the following measurement will do for the computation of their expectation values.
\begin{align} \nonumber
\Qcircuit @C=0.5cm @R=.3cm {
\lstick{n\text{=0}}& \qw &  \qw &\qw
\\
\lstick{n\text{=1}}&\ctrl{1} &\gate{H}&\meter
\\
\lstick{n\text{=2}}& \targ &  \qw & \meter
\\
\lstick{n\text{=3}}&\ctrl{1} &\gate{H}&\meter
\\
\lstick{n\text{=4}}& \targ &  \qw & \meter
}
\end{align}
Given the counts of each bit string `$b_1b_2b_3b_4$', the expectation values of $H_{XY}^{(1)}$ and $\left( H_{XY}^{(1)} \right)^2$ are respectively given by
\begin{align}
    &\left\langle H_{XY}^{(1)} \right\rangle
    =\sum_{b_1,b_2,b_3,b_4}h_{XY}^{(1)}(b_1,b_2,b_3,b_4) 
    \frac{\text{counts}_{b_1b_2b_3b_4}}{n_\text{shots}} ,    \\
    &\left\langle \left( H_{XY}^{(1)} \right)^2 \right\rangle
    =\sum_{b_1,b_2,b_3,b_4} \left( h_{XY}^{(1)}(b_1,b_2,b_3,b_4) \right)^2
    \frac{\text{counts}_{b_1b_2b_3b_4}}{n_\text{shots}} ,
\end{align}
with
\begin{align}
    h_{XY}^{(1)}(b_1,b_2,b_3,b_4) 
    :=\frac{w}{2}\sum_{n=1,3} \Bigl[ (1-2b_n)-(1-2b_n)(1-2b_{n+1}) \Bigr] .
\end{align}

For $H_Z$, the measurement in the computational basis allows us to compute
\begin{align}
    \left\langle H_{Z} \right\rangle
    &=\sum_{b_0,b_1,b_2,b_3,b_4}h_{Z}(b_0,b_1,b_2,b_3,b_4) 
    \frac{\text{counts}_{b_1b_2b_3b_4}}{n_\text{shots}}
    \\
    \left\langle  \left( H_{Z} \right)^2 \right\rangle
    &=\sum_{b_0,b_1,b_2,b_3,b_4} \left( h_{Z}(b_0,b_1,b_2,b_3,b_4) \right)^2
    \frac{\text{counts}_{b_0b_1b_2b_3b_4}}{n_\text{shots}}
\end{align}
where $h_{Z}(b_0,b_1,b_2,b_3,b_4)$ is obtained by replacing $Z_i$ with $1-2b_i$ in $H_Z$~\eqref{eq:H_decomposition-each}.
Combining these results leads to the expectation value of the total Hamiltonian~\eqref{eq:H_decomposition0},
\begin{align}
    E^{\mathrm{total}} := \langle H\rangle 
    = \left\langle H_{XY}^{(0)} \right\rangle + \left\langle H_{XY}^{(1)}\right\rangle 
      + \left\langle H_{Z} \right\rangle + C.
\end{align}
The unbiased statistical uncertainty $\delta_{\text{stat}}E^{\mathrm{total}}$ is computed as
\begin{equation}
\label{def:delta-stat}
\left( \delta_{\text{stat}}E^{\mathrm{total}} \right)^{2}
=
\frac{\left\langle \left( H_{XY}^{(0)} -\left\langle H_{XY}^{(0)} \right\rangle \right)^{2} 
  + \left( H_{XY}^{(1)} -\left\langle H_{XY}^{(1)} \right\rangle \right)^{2} 
  + \left( H_{Z} -\left\langle H_{Z}\right\rangle \right)^{2} \right\rangle}{n_\text{shots}-1}.
\end{equation}

%%%%%%%%%%%%%%%%
\subsubsection{Measurement of local energy density} 
\label{sec:measure_cond}
%%%%%%%%%%%%%%%%%%%%%%%%%

Let us move on to the discussion of local densities. 
We first consider the local behavior of the energy ($E(n)$) at each site ($n$).
Naively, one may think that its expectation value is identical to the total energy $E^{\mathrm{total}}$ divided by the volume $L$, but this is not the case for the open boundary condition 
as $E^{\mathrm{total}}/L$ is affected by the boundary.
On the other hand, if we measure the local energy density $E(n)$, then the effect of boundary is exponentially suppressed as long as the site $n$ is far from boundaries. 
Moreover, we can see how the energy density changes when we go across probe charges, so this tells more detailed information to investigate local behaviors. 

Taking into account even-odd inequality of staggered fermion, we have to take suitable average over neighboring sites to obtain the local energy density with smooth continuum limit. 
In this paper, we define the energy density at site $n$  as 
\beq
E(n) &=&  \left( \frac{h_{n-1}^M}{4} + \frac{h_n^M}{2} + \frac{h^M_{n+1}}{4} \right) +  \left( \frac{h_{n-1}^{XY}}{2} + \frac{h_n^{XY}}{2}  \right) + \left( \frac{h_{n-1}^{J}}{2} + \frac{h_n^{J}}{2}  \right),\label{eq:def-local-energy}
\eeq
where 
\beq
&& h_n^M = \frac{(-1)^n m}{2} \langle Z_n \rangle, \quad
h_n^{XY} = \frac{w}{2}\langle X_n X_{n+1} + Y_n Y_{n+1} \rangle, \nonumber\\
&& h_n^J = J q^2 \left\langle \left( \sum_{i=0}^n \frac{Z_i + (-1)^i}{2} + \frac{\theta_n}{2\pi q}       \right)^2 \right\rangle.
\eeq

%%%%%%%%%%%%%%%%
%%%%%%%%%%%%%%%%
%%%%%%%%%%%%%%%%
\section{Simulation results}
\label{sec:results}
%%%%%%%%%%%%%%%%
%%%%%%%%%%%%%%%%
%%%%%%%%%%%%%%%%
According to \eqref{eq:theo-predict-tension} and \eqref{eq:q_Schwinger_energy_small}, a negative string tension can appear in the charge-$q$ Schwinger model because of the $\mathbb{Z}_q$ $1$-form symmetry if $q>1$.
Here, we numerically investigate these properties in detail.
As simulation parameters, we take the dynamical charge $q=3$ with probe charges $q_p=-1$ and $q_p=2$, and see how the string tension changes as a function of $\theta_0$.
We also see how the $\mathbb{Z}_{q}$ 1-form symmetry emerges in the comparison of the potential and local quantities between the $q_p=-1$ and $q_p=2$ cases.

In numerical simulations, 
we implement all the operators using combinations of quantum elementary gates~\eqref{eq:gate_H}-\eqref{eq:gate_CNOT} provided by IBM Qiskit library.
Here, we consider a fixed volume simulation $gL=ga(N-1)=9.6$ and take the lattice volume $N=17$--$25$ and set $g=1.0$. 
We take the mass of the dynamical fermion in the range of $m=0.05 - 0.25$.
As for the parameters of the adiabatic preparation, we take the Trotter step, adiabatic time and initial mass to be $\delta t =0.3$, $T=99$ -- $198$ and $m_0=0.35$ -- $0.40$, respectively.
The number of shots in the measurement process is $1$ million ($n_\mathrm{shots}=10^6$), and the typical size of statistical error for $\langle H\rangle$ is $\mathcal{O} (10^{-3})\%$ for $1$ million shots. 

%%%%%%%%%%%%%%%%
%%%%%%%%%%%%%%%%
\subsection{Emergence of negative string tension}
%%%%%%%%%%%%%%%%
%%%%%%%%%%%%%%%%
%%%%%%%%%%%%%%%%
\begin{figure}[t]
\centering
\includegraphics[scale=0.3]{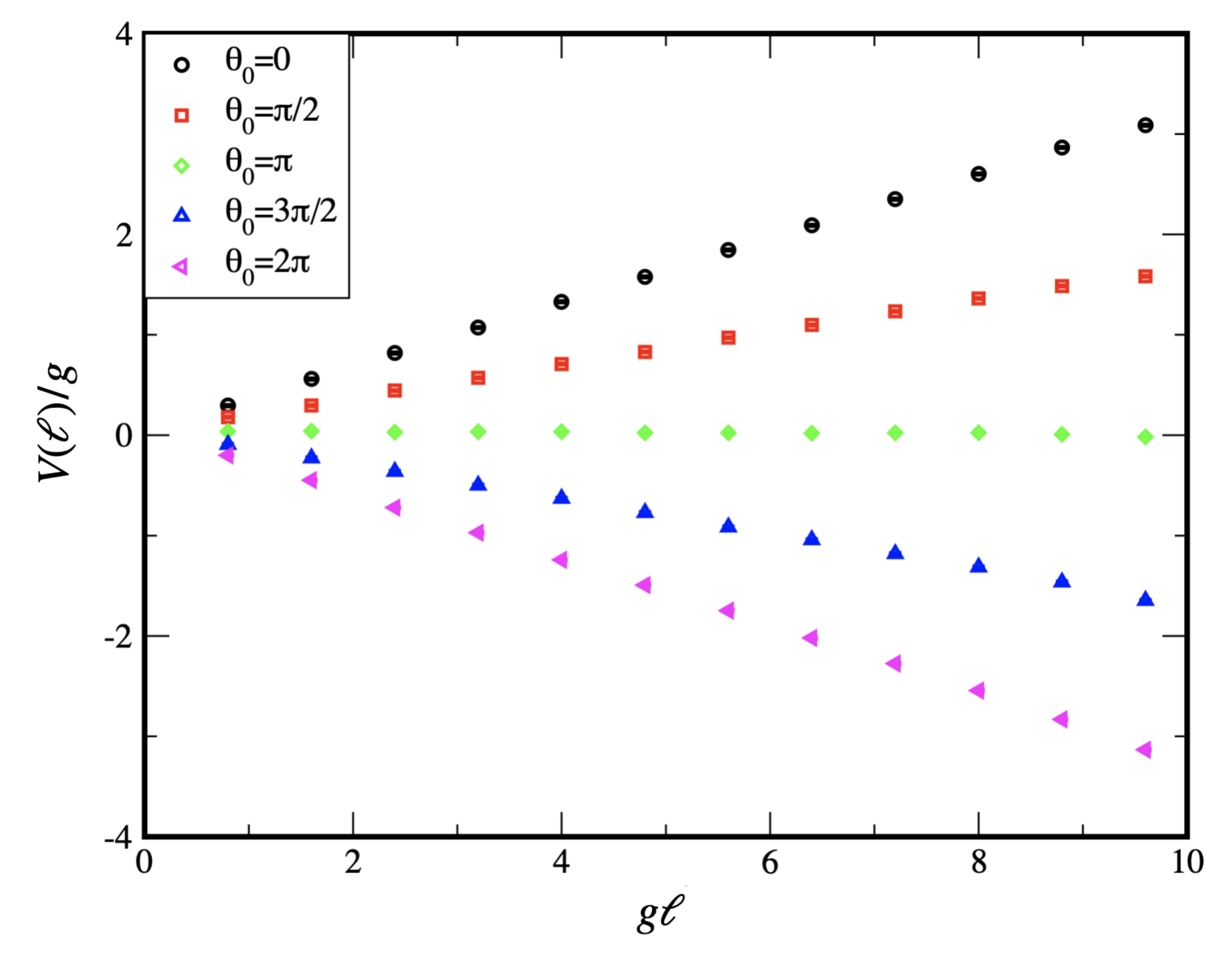}
\caption{%
$\theta_0$ dependence of the potential $V(\ell)/g$ between the two probe charges
in the charge-$q$ Schwinger model with $q=3$. 
Here we take $N=25$, $ga=0.40$, $m=0.15$ and $q_p=-1$. 
The error bars denote statistical errors.
}
\label{fig:theta0-ell-deps}
\end{figure}
%%%%%%%%%%%%%%%%

We first demonstrate that
the slope of the potential between the probes can change
its sign as varying the parameters.
Figure~\ref{fig:theta0-ell-deps} shows 
the potential $V(\ell)/g$ as a function of $g\ell$ for several values of $\theta_0$. 
Here, we take $q=3$, $m=0.15$ and $q_p=-1$.
We can see the clear linear potentials where the string tension is positive in $0 \le \theta_0 < \pi$ while it is negative in $\pi < \theta_0 \le 2\pi$. 
At $\theta=\pi$, the potential is consistent with zero in all $g\ell$ and it indicates the screening property.

%%%%%%%%%%%%%%%%
%\begin{figure}[tbp]
\begin{figure}[t]
\centering
\includegraphics[scale=0.5]{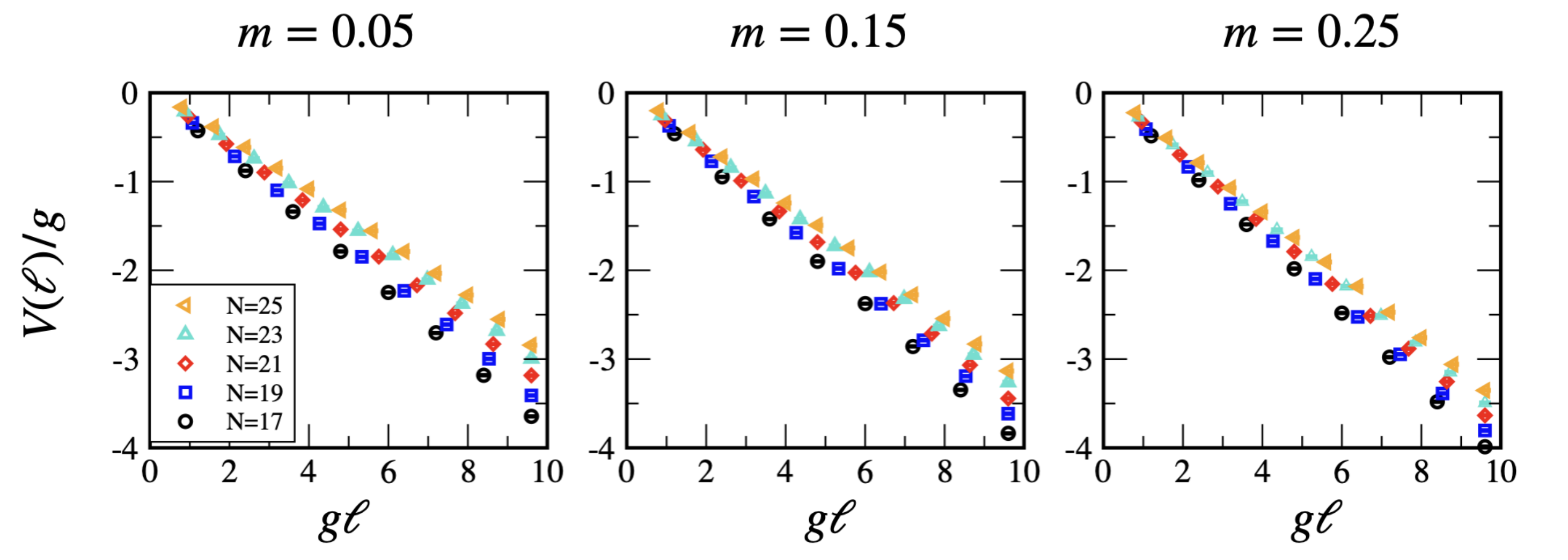}
\caption{%
The potentials $V(\ell)/g$ in the charge-3 Schwinger model for $q_p=-1$ and $\theta_0=2\pi$ are plotted for various values of $(a,N)$ with a fixed physical volume $(ga(N-1)=9.6)$.
}
\label{fig:theta0-2pi-N-deps}
\end{figure}
%%%%%%%%%%%%%%%%
%%%%%%%%%%%%%%%%%%%%%%%%%%%%
%%% string tension %%%
%\begin{table}[tbp]
\begin{table}[t]
\centering
\begin{tabular}{cc|ccc}
  \hline \hline 
  $N$ & $ga$ & $\sigma (m=0.05)/g^2$ & $\sigma (m=0.15)/g^2$ & $\sigma (m=0.25)/g^2$ \\
  \hline
$17$ & $0.60000$ & -0.380(3) & -0.397(3) &  -0.416(3) \\
$19$ & $0.53333$ & -0.354(2) & -0.377(2) &  -0.399(2) \\      
$21$ & $0.48000$ & -0.332(3) & -0.359(3) &  -0.380(3)\\    
$23$ & $0.43636$ & -0.311(2) & -0.341(3) & -0.367(2)\\    
$25$ & $0.40000$ & -0.294(3) & -0.325(3) & -0.348(3)\\
  \hline \hline
\end{tabular}
\caption{
Values of the string tensions $\sigma /g^2$
for 
$q_p=-1$ and $\theta_0=2\pi$ with the physical volume $ga(N-1)=9.6$
obtained by the fitting of the potentials in Fig.~\ref{fig:theta0-2pi-N-deps}.
}
\label{table:string-tension}
\end{table}
%%%%%%%%%%%%%%%%%%%%%%

Next we focus on the negative string tension case and 
compare with the continuum results.
Here we take $\theta_0=2\pi$ as a representative.
In Figure~\ref{fig:theta0-2pi-N-deps},
we plot the potential $V(\ell )/g$ against the probe distance $g\ell$
for various values of $(a,N)$ keeping the physical volume as $ga(N-1)=9.6$
in order to find the string tension for various values of $a$ at the fixed physical volume. 
To obtain 
the string tension ($\sigma$), 
we fit the potential for each $(a,N)$ and mass using the linear function of $\ell$ as $V(\ell) = \sigma \ell +c_0$.
We take fitting range as $3.0\le  g\ell \le 7.0$
to remove the boundary effect on finite lattice extent with open boundary condition.
Consequently, we find that the $\chi^2/$d.o.f. for all the fitting processes are reasonable, namely $\chi^2/$d.o.f. $\approx 1$.
We summarize the obtained values of the string tension by the fitting in Table~\ref{table:string-tension}.

%%%%%%%%%%%%%%%%%
%\begin{figure}[tbp]
\begin{figure}[t]
\centering
\includegraphics[scale=0.5]{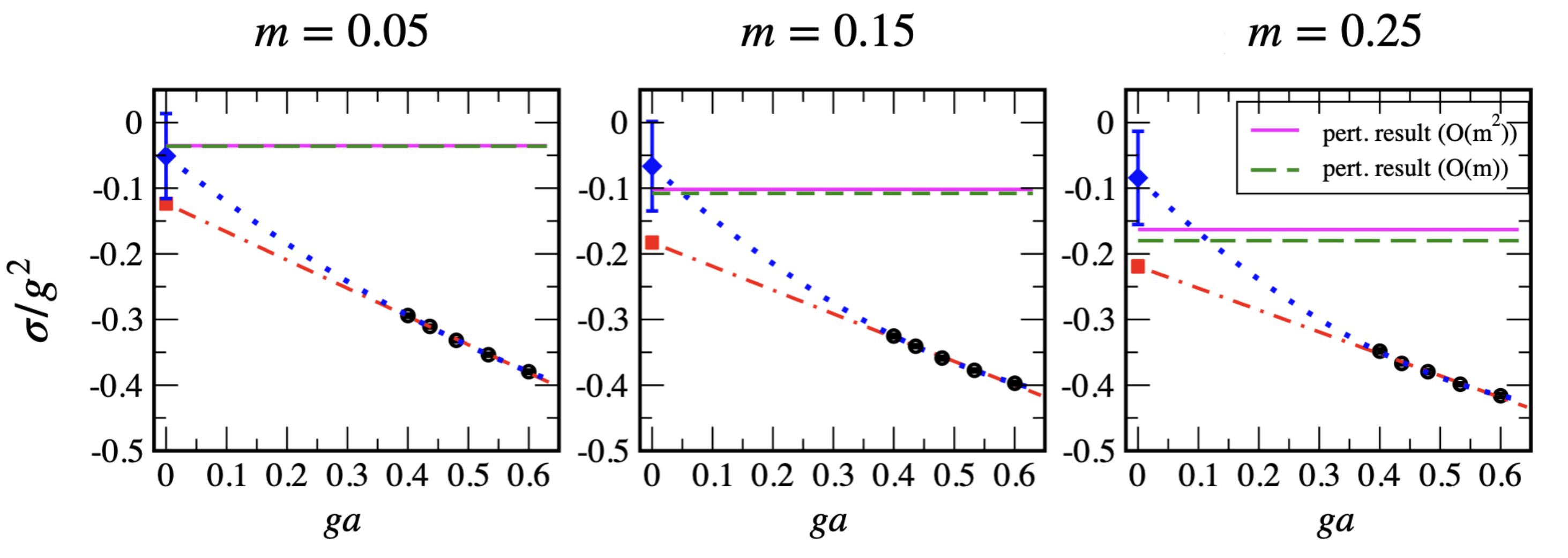}
\caption{%
Continuum extrapolations of the string tensions $\sigma /g^2$ in the charge-$3$ Schwinger model
for $q_p=-1$ and $\theta_0=2\pi$ at the physical volume $ga(N-1)=9.6$.
The linear and quadratic fits in terms of $ga$ are considered (shown with red dot-dashed lines and blue dotted curves, respectively).
The theoretical predictions in the continuum limit are based on 
the first (green dashed line) and second (magenta solid line) order approximation of the mass perturbation theory. 
The error bars associated with the black circles denote
the fitting errors in finding the sting tensions from Figure~\ref{fig:theta0-2pi-N-deps}.
}
\label{fig:cont-lim}
\end{figure}
%%%%%%%%%%%%%%%%

In terms of the values of $\sigma$ in Table~\ref{table:string-tension},
we make the continuum extrapolation for $m=0.05,0.15$ and $0.25$ as shown in Figure~\ref{fig:cont-lim}.
All the fit qualities, namely the $\chi^2/$d.o.f., in both the linear and quadratic extrapolations are reasonable.
Now we compare the obtained values of the negative string tension with the analytic predictions given by the mass perturbation theory up to $\mathcal{O}(m^2)$ in infinite volume.
The energy density of the second-order mass perturbation theory is given by\footnote{
More precisely, 
Ref.~\cite{Adam:1997wt} computed the energy density for the $q=1$ case,
which was given in Eq.~(68).
One can find the result for general $q$
by making the replacement $g \rightarrow qg$, $\theta \rightarrow (\theta - 2\pi k )/q$ in the one for $q=1$.
} \cite{Adam:1997wt}
\begin{\eq}
 E_k  (\theta_0 ) 
=-m\frac{e^\gamma qg}{2\pi^{3/2}} \cos{\frac{\theta_0 -2\pi k}{q}}
+m^2 \frac{ e^{2\gamma}}{16\pi^2} 
\left( C_+ \cos{\frac{2(\theta_0 -2\pi k )}{q}} +C_- \right)
+\mathcal{O}(m^3 ) ,
\label{eq:Adams}
\end{\eq}
where\footnote{
The precise definitions of $C_+$ and $C_-$ 
(denoted as $\mu^2 E_+ $ and $\mu^2 E_- $ in \cite{Adam:1997wt} respectively) 
are
$C_+ $ $=$ $2\pi \int_0^\infty dr \Bigl[ r \left( e^{-2K_0 (r) } -1\right) \Bigr] $
and $C_- $ $=$ $4\pi \int_0^\infty dr \Bigl[ r \log{r} \left( 
( r K_1 (r) -1 ) e^{2K_0 (r) }  +1 \right) \Bigr] $.
} 
$C_+ \simeq -8.9139$ and $C_- \simeq 9.7384$.
In the small mass cases ($m=0.05$ and $m=0.15$), 
the extrapolated values by the quadratic extrapolation in terms of $a$ are consistent with the analytic predictions 
by the mass perturbation theory in the continuum limit.
In the larger mass case ($m=0.25$), 
the quadratic extrapolation starts to deviate from the mass perturbation results.
One possibility for the deviation is that
the approximations by the mass perturbation are no longer reliable in this regime.
Indeed, at $m=0.25$, we see the non-negligible difference between
the $\mathcal{O}(m)$ and $\mathcal{O}(m^2)$ results
which may suggests importance of higher order terms in the mass perturbation theory.
The other possibility is systematic errors in our simulation.
If we include the systematic errors coming from the fitting ansatz of the extrapolation, however, then the extrapolated value is consistent, so we cannot say much within this numerical setup. 
The main source of the large systematic error is due to the smallness of $N$ in the present simulation.
If we perform a simulation for large $N$,
say $N\approx 100$, $ga \approx 0.1$ in the near future, 
then we can numerically obtain the string tension in such a large mass regime using this simulation strategy, and we may obtain clearer signal that goes beyond the mass perturbative regime.

%%%%%%%%%%%%%%%%%%%%%%%%%%%%
%%%%%%%%%%%%%%%%%%%%%%%%%%%%
\subsection{Check of $\mathbb{Z}_{q}$ $1$-form symmetry}
%\subsection{$\mathbb{Z}_{q}$ $1$-form symmetry }
\label{sec:check-symmetry}
%%%%%%%%%%%%%%%%%%%%%%%%%%%%
%%%%%%%%%%%%%%%%%%%%%%%%%%%%
%%%%%%%%%%%%%%%%
%\begin{figure}[tbp]
\begin{figure}[t]
\centering
\includegraphics[scale=0.5]{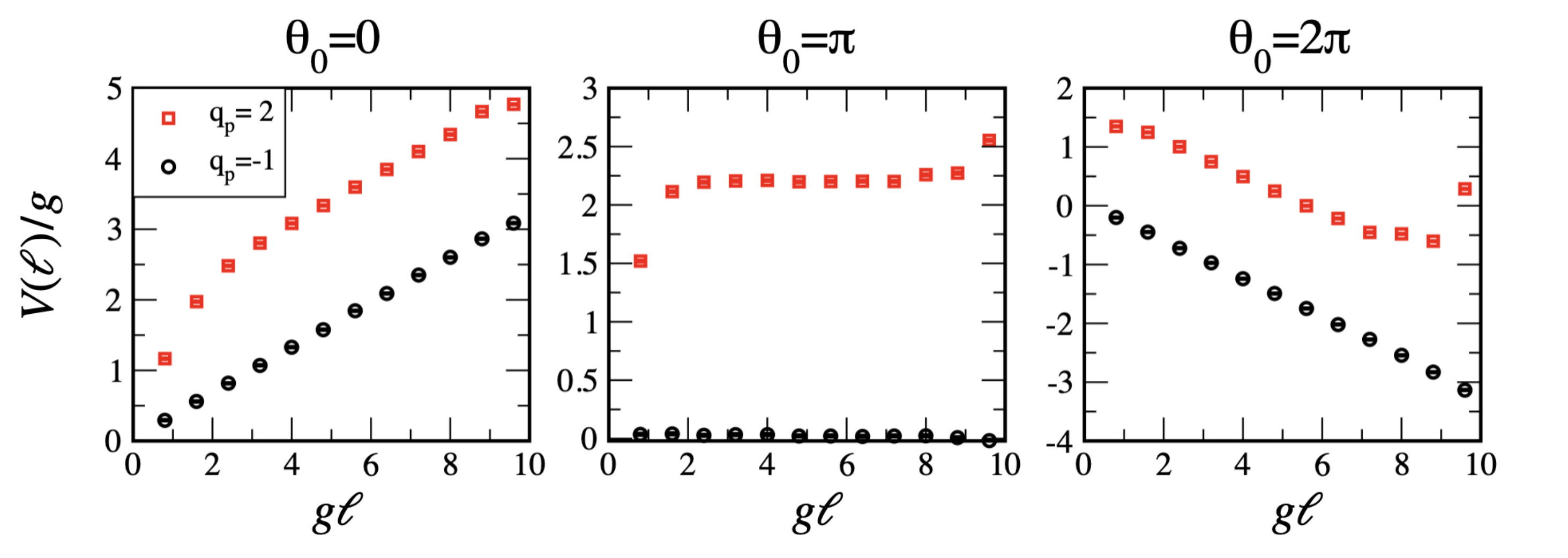}
\caption{%
Comparison of the potentials $V(\ell )/g$ between $q_p = -1$ (circle-black symbol) and $q_p=2$ (square-red symbol) 
in the charge-$3$ Schwinger model for $N=25$, $ga=0.40$ and $m=0.15$
at various values of $\theta_0$.
}
\label{fig:Comp-potential}
\end{figure}
%%%%%%%%%%%%%%%%
%%%%%%%%%%%%%%%%%%%%%%%%%%%%
%%% string tension %%%
%\begin{table}[tbp]
\begin{table}[t]
\centering
\begin{tabular}{c|c|cccc}
  \hline \hline 
 Fit range $g\ell \in [4,6]$ & $q_p$ & $\sigma (\theta_0=0 )/g^2$  &  $\sigma (\theta_0=\pi )/g^2$  &  $\sigma (\theta_0=2\pi )/g^2$   \\   
  \hline
&$2$ & $0.322(3)$ & $-0.006(5)$ & $-0.310(2)$ \\
&$-1$ & $0.322(7)$ & $-0.007(4)$ & $-0.316(2)$ \\
  \hline \hline
   Fit range $g\ell \in [3,7]$ & $q_p$ & $\sigma (\theta_0=0 )/g^2$  &  $\sigma (\theta_0=\pi )/g^2$  &  $\sigma (\theta_0=2\pi )/g^2$  \\   
  \hline
&$2$ & $0.325(3)$ & $-0.002(2)$ & $-0.303(5)$ \\
&$-1$ & $0.319(2)$ & $-0.005(1)$ & $-0.325(3)$ \\
  \hline \hline
\end{tabular}
\caption{
The string tensions $\sigma /g^2$ for 
%$q=3$, 
$N=25$, $ga=0.40$ and $m=0.15$
obtained by the fitting of the potentials shown in Fig.~\ref{fig:Comp-potential}. 
%The string tensions for $q_p=-1$ and $q_p=2$ roughly show the same value as we can expect by the $\mathbb{Z}_{q=3}$ $1$-form symmetry. 
}
\label{table:string-tension-Z3}
\end{table}
%%%%%%%%%%%%%%%%%%%%%%

Now, let us confirm the transformation property under the $\mathbb{Z}_{q}$ 1-form symmetry in our numerical simulation. 
In the previous subsection, 
we observe the stability of the negative string tension, which suggests the existence of the unconventional selection rule, 
and it is natural to interpret it as the consequence of the $1$-form symmetry. 
Here, we would like to see if the $1$-form symmetry group is actually $\mathbb{Z}_q$. 
To this end, we carry out simulations with $q_p=2$ and compare the results with $q_p=-1$. 
Since the dynamical charge is set to $q=3$, 
the two probe charges $q_p=-1$ and $2$ are the same mod $q=3$ and
therefore in the same sector of the $\mathbb{Z}_3$ $1$-form symmetry. 
This naturally predicts that they should have the same string tension
unless unknown selection rule exists. 
In numerical simulations, 
we take the longer adiabatic time $T=198$ for $q_p=2$, 
because we find that the adiabatic error for $q_p=2$ simulation is larger than the one for $q_p=-1$
(See Appendix~\ref{sec:Ad-fn} for details).
In this subsection, the mass is always set to $m=0.15$.

Figure~\ref{fig:Comp-potential} shows the comparison of the potentials between $q_p=-1$ (circle -black) and $q_p=2$ (square-red) 
for representative values of the positive ($\theta_0=0$, left panel), zero ($\theta_0=\pi$, middle panel) and negative ($\theta_0=2\pi$, right panel) string tension cases.
Clearly, we can see the slope for $q_p=2$ is similar to the one for $q_p=-1$ in the large $g\ell$ regime
if we remove a few data points near the boundary of the lattice extent.
The string tension obtained by the linear fit in terms of $\ell$ is summarized in Table~\ref{table:string-tension-Z3}.
We can see a good agreement between $q_p=-1$ and $q_p=2$ 
as expected from the $\mathbb{Z}_{q=3}$ $1$-form symmetry
although it depends more or less on the fitting range. 

%%%%%%%%%%%%%%%%
%\begin{figure}[tbp]
\begin{figure}[t]
\centering
\includegraphics[scale=0.5]{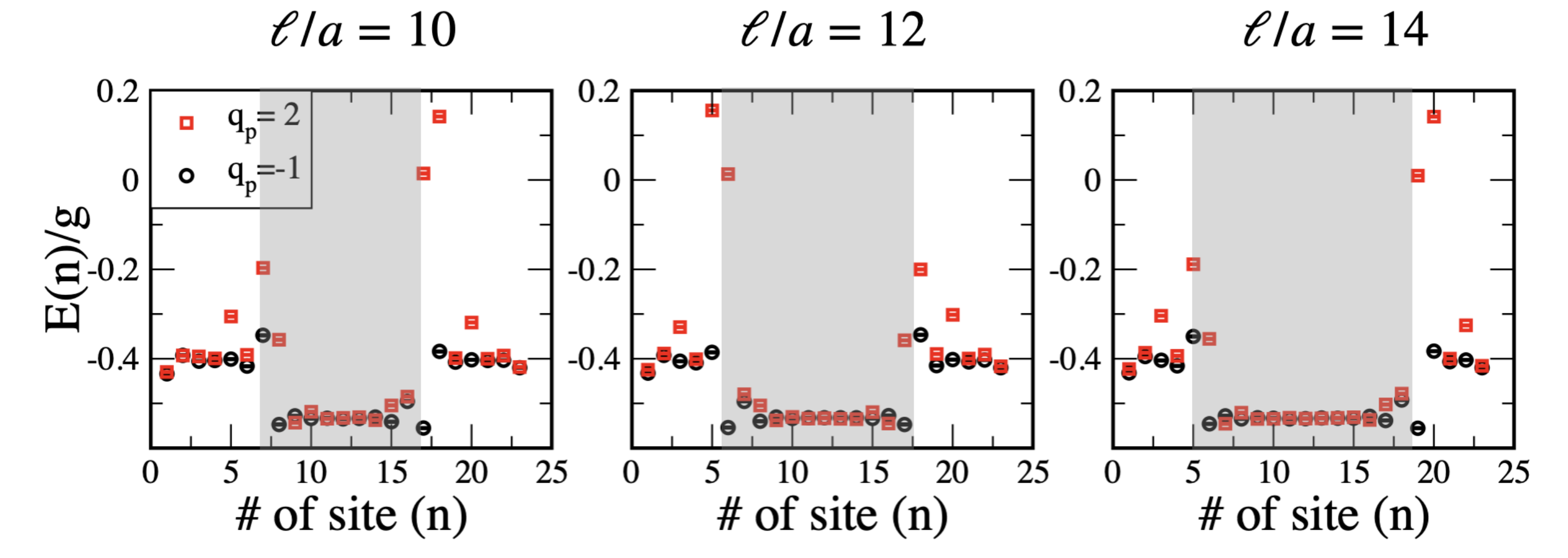}
\caption{%
Comparison of the local energy between $q_p=2$ (square-red symbol) and $q_p=-1$ (circle-black symbol)
in the charge-$3$ Schwinger model for  
$N=25$, $ga=0.40$, $m=0.15$ and $\theta_0 =2\pi$. 
The Wilson loop is extended in the shadow regime. 
}
\label{fig:local-energy}
\end{figure}
%%%%%%%%%%%%%%%%

Lastly, we study what causes the difference of the potential $V(\ell)/g$ itself between $q_p=-1$ and $q_p=2$.
To see this, we investigate the local energy density of each site.
Figure~\ref{fig:local-energy} shows 
the local energy density $E(n)/g$ defined in Eq.~\eqref{eq:def-local-energy} for $q_p=-1$ (circle-black) and $q_p=2$ (square-red) at each site in the case of $\ell/a=10,12,14$. 
The gray regime in each panel depicts the regime 
for inside of the insertion points of the probe charges or equivalently the Wilson loop .
We can see that $E(n)/g$ with $q_p=-1$ and $q_p=2$ are almost consistent with each other for inside and outside the Wilson loop, respectively. 
Thus, the discrepancy of the potential between $q_p=-1$ and $q_p=2$ comes only from the boundary contributions of the Wilson loop. 
Moreover, Figure~\ref{fig:local-energy} clearly shows that 
the energy density inside the Wilson loop is lower than the outside one, 
as suggested by the negative string tension. 
For $q_p=-1$, the energy excess at the probe charges is not so large. 
For $q_p=2$, on the other hand, the energy excess at the probe charges becomes much bigger. 
We see that 
the difference of the probe charges is localized near the insertion points, 
and the local energies for $q_p=-1$ and $q_p=2$ give the same value in the bulk. 
Because of this difference of the localized energy at the probe charges, 
the $q_p=2$ case has larger offset for the linear potential compared with the $q_p=-1$ case. 
If we remove the boundary effects, 
then it confirms the $\mathbb{Z}_{q=3}$ selection rule as predicted by the $1$-form symmetry on the lattice.

Let us discuss more detailed structure of the site-dependent energy density. 
In the limit of the massless fermion, the mass gap is given by $\mu=qg/\sqrt{\pi}\simeq 1.7g$. 
Since the fermion mass makes the mass gap larger, 
the correlation length $\xi$ can be estimated as 
\begin{equation}
    \xi\lesssim \mu^{-1}\simeq \frac{0.6}{g} . 
\end{equation}
The lattice constant for $N=25$ is set to $ga=0.40$, 
and it suggests that only a few sites around the edges of the Wilson loop are affected by the boundary effect. 
It is roughly consistent with Fig.~\ref{fig:local-energy}.

%%%%%%%%%%%%%%%%
%%%%%%%%%%%%%%%%
%%%%%%%%%%%%%%%%
\section{Summary and discussion}
\label{sec:discussion}
%%%%%%%%%%%%%%%%
%%%%%%%%%%%%%%%%
%%%%%%%%%%%%%%%%

Recent generalization of symmetries are providing us with new systematic understandings of nonperturbative QFTs, as those kinematical constraints are often realized quite nontrivially in the actual strongly correlated phenomena. 
However, these formal developments of QFTs are usually presented in the language of the path integral formulation, not in the Hamiltonian formalism. 
If we consider to simulate the strongly-coupled QFTs with quantum computers, 
then the most natural language to be used is the Hilbert space and quantum operations acting on it. 
Thus, it is an important task to understand various aspects of QFTs in the Hamiltonian formalism. 

In this paper, we considered the charge-$q$ Schwinger model on the open boundary condition, and designed the strategy to understand its properties related to the $\mathbb{Z}_q$ $1$-form symmetry using digital quantum simulation. 
When we take the periodic boundary condition, this model have the $\mathbb{Z}_q$ $1$-form symmetry and the $\mathbb{Z}_q$ chiral symmetry with a mixed 't~Hooft anomaly, so there has to be $q$ degenerate gapped vacua in the massless limit. 
In order to map the lattice Hamiltonian to the spin system for quantum computations, the local Hilbert space has to be finite dimensional. 
To achieve the criterion, 
it is convenient to take the open boundary condition, 
but some of the above features, such as $q$ degeneracy or $2\pi$ periodicity of $\theta$ angle, are lost. 
Still, there are interesting features as a remnant of $\mathbb{Z}_q$ $1$-form symmetry, 
such as the stability of negative string tensions, $\mathbb{Z}_q$ selection rule for the string tensions, nontrivial commutations between the Wilson loop and the chiral condensates, and so on.  
In particular, we paid special attention to the negative string tension in this paper to confirm the prediction of the $\mathbb{Z}_q$ $1$-form symmetry, 
as it would be one of the most exotic features of $1$-form symmetry in $(1+1)$d QFTs. 

It would be an interesting future study to look at the behavior of local chiral condensates in the presence of the probe charges. 
One of the most important features of the massless charge-$q$ Schwinger model 
is the mixed 't~Hooft anomaly between the $\mathbb{Z}_q^{[1]}$ and $(\mathbb{Z}_q)_{\mathrm{chiral}}$ symmetries, and the anomaly matching concludes $q$ degenerate vacua on closed space. 
In order to have a nontrivial check of the 't~Hooft anomaly in the Hamiltonian formalism, 
we need to find the Wilson--'t~Hooft type commutation relation between the Wilson loop and chiral condensate operators~\cite{Anber:2018jdf, Anber:2018xek, Armoni:2018bga, Misumi:2019dwq}. 
In order to observe such a commutation relation, we have to see how the condensates jump across the probe charge in the massless limit of the theory, so the use of position-dependent condensates is crucial for this purpose. 
We leave the detailed study of chiral condensates and 't~Hooft anomaly as a future work.

From quantum algorithmic perspective, 
our strategy based on the adiabatic state preparation is promising especially in the early stage of fault-tolerant era.
This is because the logical errors are expected to be well controlled at the cost of large number of physical qubits, resulting in a relatively small number of available logical qubits. 
Our careful study indeed implies that interesting physical properties that are likely to be intractable by classical computers can be extracted with the limited number of (logical) qubits.
While extensions to higher dimensional and more general gauge theories are desirable to be addressed in the future studies,\footnote{
The D-theory approach looks promising for such extensions~\cite{Chandrasekharan:1996ih, Beard:2004jr, Wiese:2021exc}. 
} 
we believe that our method presented here provides a key ingredient for such works.

%%%%%%%%%%%%%%%%%%%%%%%
%%%%%%%%%%%%%%%%%%%%%%%
\subsection*{Acknowledgment}
%%%%%%%%%%%%%%%%%%%%%%%
%%%%%%%%%%%%%%%%%%%%%%%
M.~H. is supported by MEXT Q-LEAP and JST PRESTO Grant Number JPMJPR2117, Japan.
The work of E.~I. is supported by JSPS KAKENHI with Grant Numbers 
19K03875 and JP18H05407, JST PRESTO Grant Number JPMJPR2113 and the HPCI-JHPCN System Research Project (Project ID: jh210016).
M.~H. and E.~I. are supported by JSPS Grant-in-Aid for Transformative Research Areas (A) JP21H05190.
Y.~K. is supported by the U.S. Department of Energy, Office of Science, National Quantum Information Science Research Centers, Co-design Center for Quantum Advantage, under the contract DE-SC0012704.
The work of Y.~T. is partially supported by JSPS
KAKENHI Grant-in-Aid for Research Activity Start-up,
20K22350.

%%%%%%%%%%%%%%%%%%%%%%%
%%%%%%%%%%%%%%%%%%%%%%%
%%%%%%%%%%%%%%%%%%%%%%%
\appendix
\section{Two-dimensional Maxwell theory and $U(1)$ $1$-form symmetry}
\label{sec:2dMaxwellTheory}
%%%%%%%%%%%%%%%%%%%%%%%%%%%
%%%%%%%%%%%%%%%%%%%%%%%%%%%
%%%%%%%%%%%%%%%%%%%%%%%%%%%

In this appendix, let us discuss the $1$-form symmetry in the $2$d pure Maxwell theory. 
This is a supplemental material for Sec.~\ref{sec:continuum} to explain theoretical backgrounds using the simpler model. 
Compared with the charge-$q$ Schwinger model, 
this theory has a larger $1$-form symmetry, $U(1)^{[1]}$
whose conservation law is equivalent to the equation of motion. 
In order to understand properties of $2$d $U(1)$ gauge theories from various perspectives, we perform the Euclidean path-integral quantization, the canonical quantization on $S^1$, and the canonical quantization on the interval $[0,L]$. 
This is a useful exercise to understand the $1$-form symmetry in the operator formalism.

%%%%%%%%%%%%%%%%%%%%%%%
%%%%%%%%%%%%%%%%%%%%%%%
\subsection{Euclidean theory on $\Sigma=T^2$}
\label{sec:Maxwell_Euclid}
%%%%%%%%%%%%%%%%%%%%%%%
%%%%%%%%%%%%%%%%%%%%%%%

We first discuss the $(1+1)$-dimensional pure Maxwell theory in the path-integral formalism. 
The Euclidean action is 
\be
S
=\frac{1}{2g^2}\int F\wedge \star F-\frac{\im \theta}{2\pi}\int F
=\int\diff^2 x \left( \frac{1}{2g^2}F_{12}^2 
-\frac{\im \theta}{2\pi}F_{12}\right). 
\ee
The classical equation of motion gives 
\be
\p_\mu F_{\mu\nu}=0, 
\ee
and thus $F_{12} $ must be constant. 
Let us assume that the Euclidean spacetime is a $2$-torus, $T^2=S^1_L\times S^1_T$. 
Due to the restriction of Dirac quantization, we obtain 
\be
F={2\pi n\over T L}\diff x^1\wedge \diff x^2,
\ee
for some $n\in \mathbb{Z}$, and this label $n$ refers the topological charge. The classical action for this field configuration is given as 
\be
S_n={2\pi^2\over g^2 TL}n^2-\im \theta n. 
\ee
Using Poisson summation formula, we find 
\be
Z
\propto\sum_{n}\exp(-S_n)
%\nonumber\\
\propto \sum_{k}\exp\left(-{g^2 TL\over 2}\left(k-{\theta\over 2\pi}\right)^2\right). 
\ee
As a result, we find the energy spectrum as 
\be
E_k(\theta)={g^2\over 2}\left(k-{\theta\over 2\pi}\right)^2,
\label{eq:2dMaxwell_energy}
\ee
where $k\in \mathbb{Z}$ labels the eigenstates. 

Let us revisit this result from the view point of the $U(1)$ $1$-form symmetry, denoted as $U(1)^{[1]}$. 
The Noether charge is given by 
\begin{equation}
j(x)={\im \over g^2}F_{12}+{\theta\over 2\pi},
\label{eq:U(1)_1form_Euclid}
\end{equation}
and the conservation law is given by the equation of motion,
\be 
\p_{\mu} j={\im\over g^2} \p_\mu F_{12}=0.
\ee
This means that $j$ defines a topological, point-like operator 
in the sense that the correlation functions including $j(x)$ do not depend on $x$. 
As a result, this operator $j$ does not act on any gauge-invariant local operator. 
However, as we shall see in the operator formalism, 
this has a nontrivial commutator with the extended object, called Wilson loop, $W(C)=\exp\left(\im \int_C A\right)$. 
That is, the $1$-form symmetry is a symmetry transformation acting on the test electric charge. 

We can promote this $U(1)^{[1]}$ symmetry to the local gauge redundancy, and this gives a simpler derivation for the energy eigenvalues. 
The corresponding gauge field is a $U(1)$ $2$-form gauge field $\mathcal{B}$, and the gauged action is given as 
\begin{eqnarray}
S[A,\mathcal{B}]&=&{1\over 2g^2}\int |F-\mathcal{B}|^2-{\im \theta\over 2\pi}\int(F-\mathcal{B})-{\im k}\int \mathcal{B}\nonumber\\
&=&S[A]+\im \int (j(x)-k)\wedge \mathcal{B}+O(\mathcal{B}^2). 
\end{eqnarray}
The first two terms of the first line are obtained by the minimal coupling procedure, and the last one is the discrete $\theta$ term with $k\in \mathbb{Z}$. 
The second line is the expansion in terms of $B$ and it suggests that $\mathcal{B}$ appears as an auxiliary field, constraining the Noether charge $j(x)=k$. 
This action is invariant under the $U(1)$ $1$-form gauge transformation,\footnote{Although the discrete theta term is not manifestly invariant, because of the Dirac quantization  $\int \diff \Lambda \in 2\pi \mathbb{Z}$, $\exp(-S)$ is invariant when $k\in \mathbb{Z}$. } 
\be 
A\mapsto A+\Lambda,\quad 
\mathcal{B}\mapsto \mathcal{B}+\diff \Lambda, 
\ee
where the gauge transformation parameter $\Lambda$ itself is a $U(1)$ gauge field. 
Here, $\Lambda$ is not necessarily a flat connection. 
Using this gauge transformation, we can set $F=0$ as a gauge-fixing condition, and then we find 
\bea
Z_k(\theta)&:=&\int \Diff \mathcal{B} \int \Diff A \exp\left(-{1\over 2g^2}\int |F-\mathcal{B}|^2+{\im \theta\over 2\pi}\int(F-\mathcal{B})+{\im k}\int \mathcal{B}\right)\nonumber\\
&=&\int \Diff' \mathcal{B}\exp\left(-{1\over 2g^2}\int |\mathcal{B}|^2+\im\left(k-{\theta\over 2\pi}\right)\int \mathcal{B}\right)\nonumber\\
&=&\exp\left(-TL E_k(\theta)\right). 
\eea
Thus, gauging of $U(1)$ $1$-form symmetry with the discrete $\theta$ term gives the projection to the $k$-th branch of ground states, $E_k(\theta)$, given in (\ref{eq:2dMaxwell_energy}).

%%%%%%%%%%%%%%%%%%%%%%%%%%%
%%%%%%%%%%%%%%%%%%%%%%%%%%%
\subsection{Canonical quantization on $S^1_L\times \mathbb{R}_t$}
%%%%%%%%%%%%%%%%%%%%%%%%%%%
%%%%%%%%%%%%%%%%%%%%%%%%%%%

Let us reproduce the result obtained by the path-integral method in the Hamiltonian formulation. 
In this subsection, we take the space manifold as $S^1_L$. 
In Minkowski formulation, the Lagrangian is 
\be
\mathcal{L}
={1\over 2g^2}F_{01}^2+{\theta\over 2\pi}F_{01} .
\ee
In the temporal gauge, the Hamiltonian density is given by
\be
H=\Pi \dot{A_1}-L={g^2\over 2\pi}\left(\Pi-{\theta\over 2\pi}\right)^2 ,
\ee
where the physical states must satisfy the Gauss law constraint
\be
\p_x \Pi=0. 
\label{eq:Gauss_law}
\ee
In the canonical quantization, we replace $\Pi$ as $-\im{\delta\over \delta A_1(x)}$. 

Now, we have obtained the Hamiltonian, so let us specify the Hilbert space. 
Let $\Psi[A_1]$ be a wave functional. In order for this to be physical, it must satisfy the Gauss law (\ref{eq:Gauss_law}), 
\be
\p_x {\delta \Psi\over \delta A_1(x)}=0,
\ee
which means that $\Psi$ in invariant under $A_1(x)\to A_1(x)+\p_x \ve(x)$ for some small $\ve(x)$, which is thus periodic on $S^1_L$ as $\mathbb{R}$-valued functions. 
Therefore, the Hilbert space should be spanned by the ``spatial Wilson loops'',
\be
\Psi_k[A_1]=\exp\left(\im k \int_0^L A_1(x)\diff x\right). 
\label{eq:eigenfunction}
\ee
At this stage, the label $k$ can be an arbitrary real number. 
By \textit{also} requiring the invariance under the large gauge transformation, $A_1(x)\to A_1(x)+{2\pi\over L}$, we obtain
\be
k\in\mathbb{Z}.
\ee
As this is the eigenvalue of $\Pi$,  
$\Pi \Psi_k = k \Psi_k,$
we obtain 
\be
E_k(\theta)={g^2\over 2}\left(k-{\theta\over 2\pi}\right)^2,  
\ee
and the path-integral result is reproduced. 
We note that the quantization of $k$ comes out of the Dirac quantization of the topological charge in the path-integral method, while it is the consequence of invariance under the large gauge transformation in the operator formalism. 

Let us rephrase these results in terms of the $U(1)^{[1]}$ symmetry. 
We note that the $1$-form symmetry generator $j(x)$ in (\ref{eq:U(1)_1form_Euclid}) becomes $\Pi(x)$ in (\ref{eq:conjugate_momentum}) by the Wick rotation. 
Let us consider the operator 
\be 
U_{\alpha}(x)=\exp\left(\im \, \alpha \Pi(x)\right),  
\ee
then we find 
\be 
U_{\alpha}(x)\Psi_k[A_1]=\rme^{\im \alpha k} \Psi_k[A_1]. 
\ee
As $\Psi_k$ is obtained by multiplying the Wilson loop along $S^1$, this result shows that $U_\alpha(x)$ and the Wilson loop $W_k(S^1)$ has a nontrivial commutation relation, and gives a phase factor $\rme^{\im \alpha k}$, where $k$ is the charge of the Wilson loop. 
We also note that $\alpha=2\pi$ corresponds to a large gauge transformation. 
As we have required the invariance of the whole states under the large gauge transformation, $U_{2\pi}(x)$ acts trivially, and we can regard $U_{2\pi}(x)=1$. 
This gives the periodicity of the transformation parameter, $\alpha\sim \alpha+2\pi$, which confirms that the $1$-form symmetry group is actually $U(1)$, not $\mathbb{R}$. 

%%%%%%%%%%%%%%%%%%%%%%%%%%%
%%%%%%%%%%%%%%%%%%%%%%%%%%%
\subsection{Canonical quantization on $[0,L]\times \mathbb{R}_t$}
\label{sec:Maxwell_canonical_interval}
%%%%%%%%%%%%%%%%%%%%%%%%%%%
%%%%%%%%%%%%%%%%%%%%%%%%%%%

As we formulate our lattice Schwinger model with the open boundary condition, 
let us also consider the open interval $[0,L]$ instead of $S^1$ as the spatial manifold. 

The Hamiltonian operator is the same as before, so the eigenfunctions take the same form~(\ref{eq:eigenfunction}).
Under the gauge transformation, $A_1(x)\mapsto A_1(x)+\p_x \lambda(x)$, 
the wave functional behaves as 
\bea
\Psi_k[A_1]&\mapsto & \exp\left(\im k \int_0^L \p_x \lambda \diff x\right)\Psi_k[A_1]\nonumber\\
&=& \exp(\im k \lambda(L)-\im k \lambda(0)) \Psi_k[A_1]. 
\eea
Thus, only when $k=0$, the ``naive'' Gauss law (\ref{eq:Gauss_law}) is satisfied. 

We note, however, that the violation of gauge invariance occurs at the boundaries, 
and that this does not mean $\Psi_k[A_1]$ is unphysical for $k\not=0$. 
By putting charge-$k$ Wilson loop on the boundary, 
the Lagrangian is affected as follows, 
\be
L={1\over 2g^2}F_{01}^2+{\theta\over 2\pi}F_{01}+k A_0(x=L)-k A_0(x=0). 
\ee
The Hamiltonian operator is not affected as we will eventually take the temporal gauge, but the Gauss law is modified as 
\be
\p_x \Pi(x)+k\left(\delta(x-L)-\delta(x)\right)=0, 
\ee
and $\Psi_k[A_1]$ satisfies this modified version of the Gauss law constraint. 
Therefore, $\Psi_k[A_1]$ is a physical ground state for the system with the test charge $k$ at the boundaries, or at infinities in the limit $L\to \infty$.  

We note that, on closed spatial manifold $S^1_L$, the $\theta$ angle shows $2\pi$ periodicity by level crossing phenomenon. 
On the other hand, the $\theta$ periodicity is completely lost for the fixed open boundary condition. 
If we \textit{would like to} recover it, 
then we must consider the distinct sectors, distinguished by the charges at the boundaries.
The $1$-form symmetry generator $\Pi$ measures those charges, as $\Pi(x) \Psi_k=k \Psi_k$. 

%%%%%%%%%%%%%%%%%%%%%%%
%%%%%%%%%%%%%%%%%%%%%%%
%%%%%%%%%%%%%%%%%%%%%%%
\section{Choice of the adiabatic schedule and adiabatic error}
\label{sec:Ad-fn}
%%%%%%%%%%%%%%%%%%%%%%%
%%%%%%%%%%%%%%%%%%%%%%%
%%%%%%%%%%%%%%%%%%%%%%%
We investigate how the adiabatic schedule ($f(s)$ in \eqref{eq:ad-fn}) affects the adiabatic error.
The adiabatic theorem guarantees that the desired ground state is obtained under the assumption that the adiabatic Hamiltonian has a unique ground state along the adiabatic path 
if the adiabatic time $T$ is taken to be infinity~\eqref{eq:adiabatic-prep}. 
However, in practice, we should take the adiabatic time $T$ to be finite and
this results in a systematic (adiabatic) error~\cite{messiah1962quantum}.

Suppose that an adiabatic Hamiltonian $H_\mathrm{A}(s)$ possesses a unique ground state for all $s\in[0,1]$. 
If we wish to prepare state $\ket{\mathrm{GS_A}}$ that approximates the ground state $\ket{\mathrm{GS}}$ of $H=H_\mathrm{A}(T)$ with the precision $\epsilon$, i.e., 
\begin{align}
    \| \ket{\mathrm{GS_A}}-\ket{\mathrm{GS}}\| \le \epsilon,
\end{align}
then the adiabatic time is roughly lower-bounded as
\begin{align}
    T \gtrsim \frac{1}{\epsilon}\max_s\frac{\frac{\diff}{\diff s} H_\mathrm{A}(s)}{|E_1(s)-E_0(s)|^2}.
\end{align}
Here, $E_0(s)$ and $E_1(s)$ are the ground and first-excited states of $H_\mathrm{A}(s)$ respectively (see, e.g.,~\cite{Jansen_2007} for more elaborated analysis of the adiabatic error.).
Hence, the adiabatic error crucially depends on the energy gap between the ground and first-excited states along the evolution as well as the adiabatic time $T$.

In the present work, we introduce the $s$ dependence of $H_A(s)$ by changing parameters of the model as follows,
\begin{align}
 &w\to w f(s), \quad
 \theta_0\to \theta_0 f(s), \quad
 q_p \to q_p f(s),  \quad
 m\to m_0 \left(1-f(s) \right) + m f(s).
\label{eq:ad-deps-para}
\end{align}
Here, we suppose that the schedule function $f(s)$ is smooth function in $s \in [0,1]$ and satisfies $f(0)=0$ and $f(1)=1$.
In this appendix, we numerically investigate how the ground state obtained with the adiabatic evolution depends on the choice of the interpolating function $f(s)$ by trying several functional forms of $f(s)$.\footnote{
In this Appendix, we use the ``snapshot" functionality of Qiskit for quantum simulation without statistical uncertainties.
The snapshot utilizes the function of quantum simulator and the corresponding calculation does not exist in a real quantum computer.
However, the data of snapshot corresponds to the average of the infinite number of shots and does not suffer from statistical fluctuations.
Therefore, it is useful to see a systematic uncertainty of quantum simulation.
}

%%%%%%%%%%%%%%%%
%\begin{figure}[htbp]
\begin{figure}[t]
\centering
\includegraphics[scale=0.5]{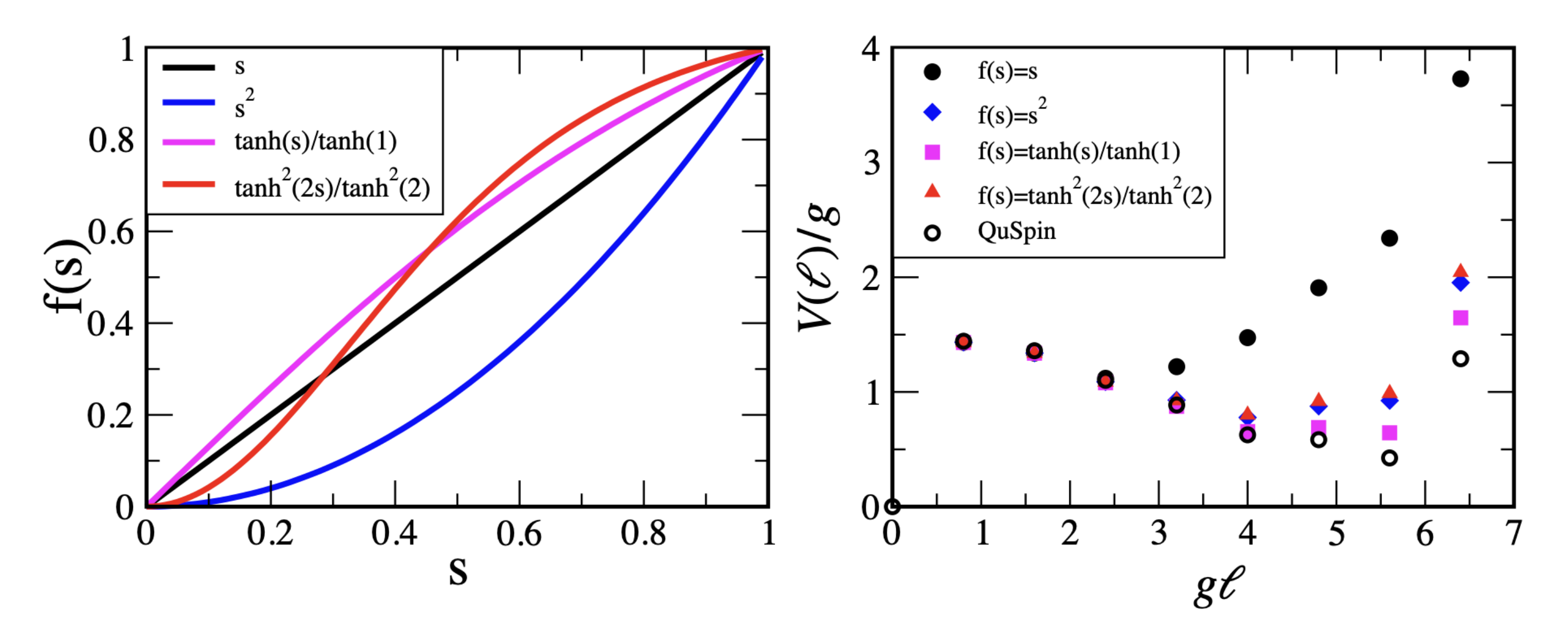}
\caption{%
Left: Several functions $f(s)$ here we investigate. 
Right: The potential values for several choices of adiabatic schedule function $f(s)$. Open-circle symbol denotes the data calculated by the Python package QuSpin (exact diagonalization)~\cite{weinberg2017quspin,weinberg2019quspin}. The other symbols denote the data with the shorter adiabatic time $T=99$ with several choice of function $f(s)$, where the color is the same with the one in left panel.
}
\label{fig:ad-fn-deps}
\end{figure}
%%%%%%%%%%%%%%%%
Figure~\ref{fig:ad-fn-deps} depicts the adiabatic schedule functions (left panel) and the corresponding data of potential (right panel).
Here, we take $N=17, ga=0.40, m=0.20, q_p=2, \theta_0=2\pi$, but we obtain similar result for other lattice size as well if we fix $q_p=2, \theta_0=2\pi$. 
In the right panel, 
the open-circle symbol denotes the data calculated by the Python package QuSpin (exact diagonalization).  
The other filled symbols denote the data obtained by an adiabatic state preparation with the adiabatic time $T=99$ using several schedule functions, 
where each color is the same with the one in left panel.
The difference between each filled symbol and open symbol at each $g\ell$ represents its adiabatic error and Trotter error. 
Here, we fix the size of the Trotter step as $\delta t=0.3$, 
and we find that the adiabatic error is dominated in this simulation setup.  
In the left panel of Fig.~\ref{fig:ad-fn-deps}, 
we can find that the data with $f(s)=\tanh(s)/\tanh(1)$ (square-magenta) has the smallest error.

\begin{figure}[t]
\centering
\includegraphics[scale=0.7]{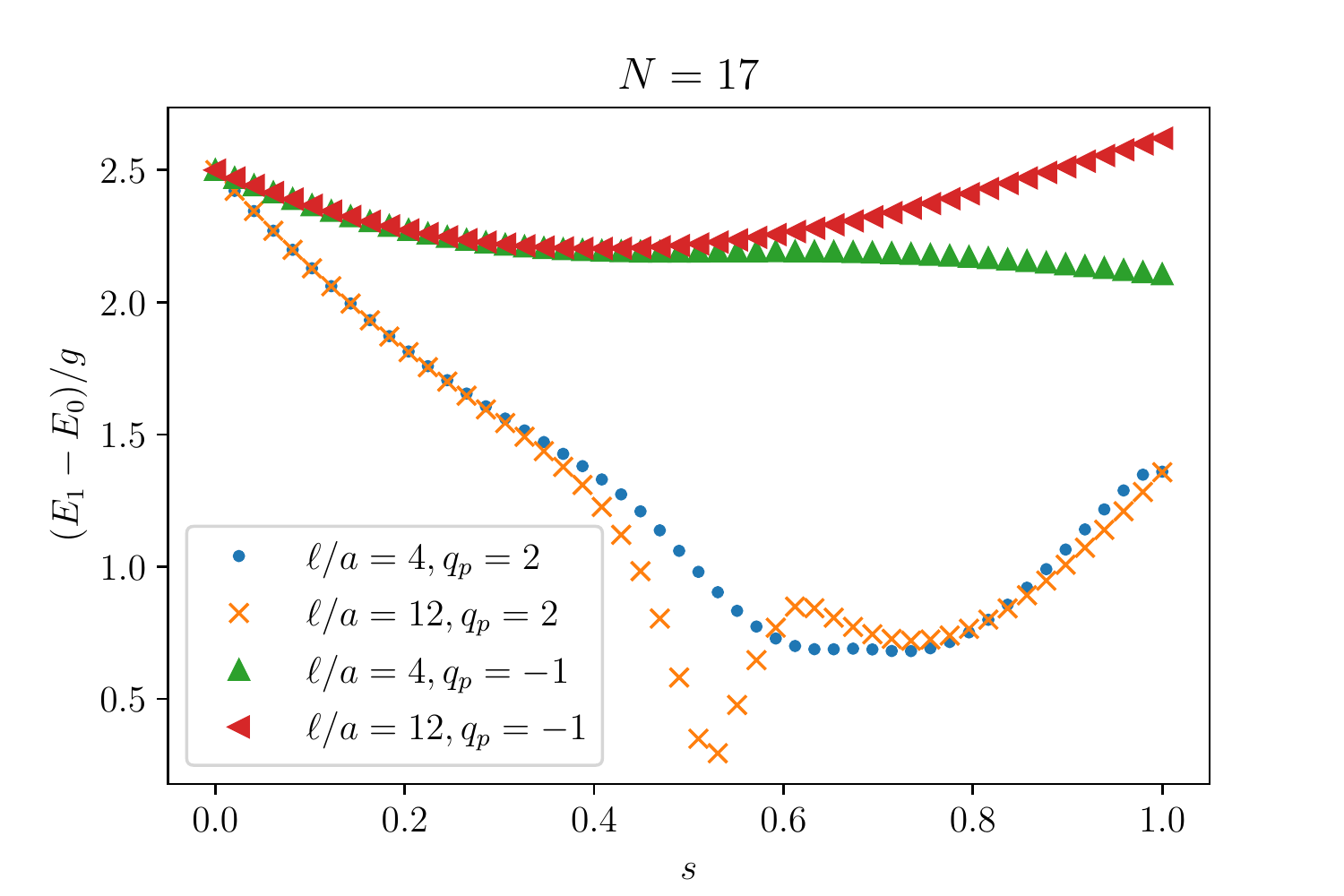}
\caption{%
Instantaneous energy eigenvalues of the ground and first excited states of the adiabatic Hamiltonian.
}
\label{fig:energy-gap}
\end{figure}

As we mentioned in Section~\ref{sec:check-symmetry},
we find the adiabatic error for $q_p=2$ is larger than the one for $q_p=-1$
and therefore we take the longer adiabatic time for $q_p =2$ in the main text.
One might expect that if we change the value of the probe charge from $q_p=2$ to $q_p=-1$, 
then we would also see a similar behavior because of the $\mathbb{Z}_{q}$ symmetry.
However, 
we could not see such a strong dependence on choice of $f(s)$ in the $q_p=-1$ case.
To investigate an origin of the difference,
we calculate the instantaneous energy gap between the ground and the first excited states of the adiabatic Hamiltonian using exact diagonalization 
as shown in Fig.~\ref{fig:energy-gap}. 
Here, we take the linear adiabatic schedule function, $f(s)=s$.
%%%%%%%%%%%%%%%%
%\begin{figure}[htbp]
%%%%%%%%%%%%%%%%
The left-triangle (red) and up-triangle (green) symbols denote $g\ell = 4$ and $g\ell=12$ in the case of $q_p=-1$, respectively.
We can see the energy gap $E(n)/g$ takes more than $2.0$ and it is almost constant during the adiabatic time evolution.
On the other hand, circle (blue) and cross (orange) symbols denote $g\ell = 4$ and $g\ell=12$ in the case of $q_p=2$, respectively.
The energy gap is smaller than the one in the case of $q_p=-1$.
It may be attributed to the evolution of total electric flux $\vartheta = \theta_0  +2\pi q_p$ in \eqref{eq:ad-deps-para}.
The total electric flux remains to be $\vartheta =0$ in the time evolution $s=0 \rightarrow s=1$ if we take $q_p=-1$ with $\theta_0=2\pi$.
On the other hand, 
it changes $\vartheta =0 \rightarrow \vartheta =6\pi $ 
if we take $q_p=2$ with $\theta_0=2\pi$.
Thus, the magnitude of the change of the total electric flux is small in the $q_p=-1$ case, and it reduces the adiabatic error.

Furthermore, we can see the energy gap of $g\ell=12$ with $q_p=2$ has a local minimum around $s=0.5$.
The other symbols also have an inflection around $s=0.5$.
It reminds us that  there is a phase transition at $\theta_0/(2\pi)=0.5$ in the $q=1$ massive Schwinger model~\cite{Coleman:1976uz, Byrnes:2002nv}.
The local minimum of energy gap may indicate the existence of a similar phase transition around $\vartheta/(2\pi q) = 0.5$, 
which is realized at $s=0.5$ in the charge-$q$ massive Schwinger model.
That expectation is related to the periodicity of $\theta_0$ in the charge-$q$ Schwinger model as discussed in \S.~\ref{sec:periodicity}, we will report it near future.

%%%%%%%%%%%%%%%%%%%%%%%
%%%%%%%%%%%%%%%%%%%%%%%
%%%%%%%%%%%%%%%%%%%%%%%
%\begin{thebibliography}{99}
\bibliographystyle{utphys}
\bibliography{quantum_computation,./QFT}
%%%%%%%%%%%%%%%%%%%%%%%
%%%%%%%%%%%%%%%%%%%%%%%
%%%%%%%%%%%%%%%%%%%%%%%
%\end{thebibliography} 

\end{document}